\documentclass[twoside,11pt]{article}

\usepackage{jmlr2e}
\usepackage{csvsimple}
\usepackage[english]{babel}
\usepackage{amsfonts}
\usepackage{amsmath}
\usepackage{amssymb}
\usepackage{hyperref}
\usepackage[latin1]{inputenc}
\usepackage{graphicx}
\usepackage{listings}
\usepackage{pgfplots}

\pgfplotsset{compat=1.13}

\begin{document}

\title{Topic Grouper: An Agglomerative Clustering Approach to Topic Modeling}

\author{\name Daniel Pfeifer \email daniel.pfeifer@hs-heilbronn.de \\
       \addr Department of Medical Informatics\\
       University of Heilbronn\\
       Max-Planck-Str. 39, 74081 Heilbronn, Germany
       \AND
       \name Jochen L.~Leidner \email jochen.leidner@refinitiv.com \\
       \addr Refinitiv Labs,\\
       30 South Colonnade, London E14 5EP, United Kingdom\\
       \addr University of Sheffield,\\
			 Department of Computer Science,\\
       211 Portobello, Sheffield S1 4DP, United Kingdom}

\editor{??}

\maketitle


\begin{abstract}
We introduce Topic Grouper as a complementary approach in the field of
probabilistic topic modeling.  Topic Grouper creates a disjunctive
partitioning of the training vocabulary in a stepwise manner such that
resulting partitions represent topics.  It is governed by a simple
generative model, where the likelihood to generate the training
documents via topics is optimized.  The algorithm starts with one-word
topics and joins two topics at every step. It therefore generates a
solution for every desired number of topics ranging between the size
of the training vocabulary and one. The process represents an
agglomerative clustering that corresponds to a binary tree of topics. A
resulting tree may act as a containment hierarchy, typically with more
general topics towards the root of tree and more specific topics
towards the leaves. Topic Grouper is \emph{not} governed by a background
distribution such as the Dirichlet and avoids hyper parameter
optimizations.

We show that Topic Grouper has reasonable predictive power and also a
reasonable theoretical and practical complexity.  Topic Grouper can
deal well with stop words and function words and tends to push them
into their own topics. Also, it can handle topic distributions, where
some topics are more frequent than others. We present typical examples
of computed topics from evaluation datasets, where topics appear 
conclusive and coherent. In this context, the fact that each word
belongs to exactly one topic is not a major limitation; in some
scenarios this can even be a genuine advantage, e.g.~a related
shopping basket analysis may aid in optimizing groupings of articles
in sales catalogs.\footnote{A shorter version of this paper has been published by Springer-Verlag GmbH, Heidelberg (see \cite{10.1007/978-3-030-15712-8_38}).} 
\end{abstract}



 
\section{Introduction}
Over the last two decades, probabilistic topic modeling (\emph{topic
  modeling} for short) has become an active sub-field of
information retrieval and machine learning.  Topic modeling may be
considered a refinement of document clustering and comes as an
unsupervised machine learning approach in its basic versions: as
opposed to pure document clustering, \emph{topic modeling allows for
  many topics to occur in a single document} but still mandates common
topics across the documents of a training collection. Hereby, each
topic is typically represented via a multinomial distribution over the
collection's vocabulary. Related ideas and solutions were formed in
the two seminal publications on \emph{probabilistic Latent Semantic
  Indexing} (pLSI) (\cite{hofmann:1999:uai}) and \emph{Latent
  Dirichlet Allocation} (LDA)
(\cite{blei:2003:lda:944919.944937}). \cite{Pritchard-Stephens-Donnelly:2000:Genetics}
proposed a model similar to LDA independently in the field of
population genetics.


Besides classical text document analysis and genetics, topic modeling has turned out to be of
use in bio-informatics \citep{topicmodelingbio:2016}, digital libraries \citep{griffiths-steyvers:2004:pnas},
recommender systems \citep{hu-hall-attenberg:2014:kdd}, computing in the service of
political and social studies (``digital humanities'') \citep{journaldh:2012} and other application
areas (e.g.~see \cite{boydgraber-hu-mimno:2017:ftir}).

Regarding pure document clustering, the two major machine learning
directions are \emph{Expectation Maximization} (EM) including $k$-Means on
the one hand and hierarchical clustering including agglomerative
clustering on the other hand.  In comparison, EM-based techniques have also
been a central means for topic inference but \emph{the opportunities of
  hierarchical clustering for topic modeling have been overlooked to date.} 
In this paper, we aim to partially close this gap by developing and
evaluating \emph{Topic Grouper} as \emph{a topic modeling approach based on agglomerative clustering}.

Important benefits of agglomerative clustering for topic modeling lie
in its simplicity, absence of hyper parameters, deep hierarchical
structures of topics as well as the ability to find even conceptually
narrow topics. A major challenge is to determine a well-founded
cluster distance with reasonable predictive qualities and
computational performance.


The remainder of this article is structured as follows: Section~\ref{sec:related-work} describes relevant related work.  
It also outlines basic concepts behind topic models and summarizes the hyper parameter problem for LDA.
Section~\ref{sec:theory} introduces the generative model behind Topic Grouper and derives a related cluster distance.
Moreover, a corresponding algorithm for model computation is presented and its complexity is assessed.
Section~\ref{sec:evaluation} includes a range of experiments comparing the performance of Topic Grouper with two LDA variants.
A synthetic dataset allows for applying error rate as a quality measure. Regarding real-world datasets covering retailing and text, we resort to perplexity.
In addition, Section~\ref{featurered} examines Topic Grouper as a feature reduction method for text classification and compares it to LDA as well as to two common text-oriented feature selection techniques.
Section~\ref{viz} discusses approaches to inspect learned models and reports on related examples for a larger text dataset.
Section~\ref{sec:discussion} summarizes and discusses our findings.
Section~\ref{sec:conclusion} gives pointers to possible future work.


\section{Basic Concepts and Related Work} \label{sec:related-work}

\subsection{Agglomerative Clustering} 

Clustering items of data, such as sets of vectors of numbers by
similarity is an old idea.  \emph{Hierarchical agglomerative
  clustering} (HAC) or simply agglomerative clustering is the process
of clustering the clusters in turn iteratively, based on a similarity
measure between clusters from a previous iteration. It was first
described in the 1960s by authors including\cite{ward:1963:jamstat},
\cite{Lance-Williams:1966:Nature,Lance-Williams:1967:ComputerJ},
and others.

A \emph{cluster distance} is usually the term for the inverse of a similarity measure underlying a clustering procedure.
Standard cluster distances derived from the so-called Lance-Williams formula 
include single linkage, complete linkage and group average linkage, but many others have been proposed (see,~\cite{Murtagh83,xu-survey-clustering-algorithms-2005}). 

Cluster distances, such as the one developed here, may not necessarily meet standard mathematical distance axioms, 
as agglomerative clustering can do without (\cite{ward:1963:jamstat}). Moreover, our cluster distance is \emph{model-based}, as it is governed by 
a simple generative model. Model-based agglomerative clustering has rarely been investigated:
\cite{Kamvar:2002:IEC:645531.656166} give a model-based interpretation of some standard cluster distances and partly extend them under the same framework.
\cite{Vaithyanathan:2000:MHC:2073946.2074016} develop a recursive probabilistic model for a clustering tree in order to explain the data items merged at each tree node.
The model is applied to the case of pure document clustering. For efficiency reasons the authors resort to a mix of agglomerative and flat clustering.

A common critique of agglomerative clustering is its relatively high time complexity typically 
amounting to $O(k^2)$ or more given the number of data items $k$ (\cite{xu-survey-clustering-algorithms-2005}). Also, space complexity is often in $O(k^2)$ depending on the chosen cluster distance.
In the case of our contribution and additionally in the case of text, \emph{$k$ corresponds to the vocabulary size}, 
\emph{which can be limited} even for large text collections, e.g.~by simple filtering criteria such as high document frequency.
This offers the potential for a reasonable computational overhead in the context of topic modeling.

A major asset of agglomerative clustering is the \emph{tree structure of its clusters} often assumed to reflect containment hierarchies.
Also, it is widely held that agglomerative clustering offers better and more computationally stable clusters 
than competing procedures such as $k$-Means (\cite{Jain88}, p.~140).

For further exposition, we refer the reader to recent text books on
the topic (e.g.~\cite{Xu-Wunsch:2008,Everitt-etal:2011}) and various survey papers (e.g.~\cite{Murtagh83}, \cite{Jain:1999:DCR:331499.331504} and \cite{xu-survey-clustering-algorithms-2005}).


\subsection{Probabilistic Topic Modeling}

\label{sec:concepts}

Topic modeling evolved from \emph{Latent Semantic Analysis} (LSA) -- an algebraic dimensionality
reduction technique using \emph{Singular Value Decomposition} to
retain the $n$ largest singular values which show the dimensions with
the greatest variance between words and documents
(\cite{deerwester-etal:1990:jasist}).  \emph{Latent Semantic Indexing} is the
application of LSA to document indexing and retrieval
(\cite{hofmann:1999:sigir}).  A drawback of LSA is the
lack of a probabilistic interpretation.  This was first addressed
by pLSI in \cite{hofmann:1999:uai}.

In their influential paper, \cite{blei:2003:lda:944919.944937}
describe LDA and extend pLSI by two Dirichlet priors, thus completing
the generative approach and aiding in smoothing of the resulting
models.  In the following, we briefly reiterate such non-hierarchical
or \emph{flat} topic models in order to provide the foundation for
 our own method.
\newpage
\subsubsection{Non-Hierarchical Topic Models}

Let
\begin{itemize}
\item $D$ be the set of training documents with size $|D|$,
\item $V$ be the vocabulary of $D$ with size $|V|$,
\item $f_d(w)$ be the frequency of a word $w \in V$ with regard to $d \in D$.
\end{itemize}
Given a set of topic references $T$ with $|T| = n$, the goal of
non-hierarchical or flat topic modeling is to estimate respectively consider $n$
topic-word distributions $p(w|t)_{w \in V}$ (one for each $t \in T$) and
$|D|$ document-topic distributions $p(t|d)_{t \in T}$ (one for each $d \in
D$).  Together, these distributions are meant to maximize $p(D) =
\prod_{d \in D} p(d)$, where $p(d)$ is the probability of all word
occurrences in $d$ regardless of their order.  Yet, how this is done
in detail, depends on the topic modeling approach: Under pLSI
(\cite{hofmann:1999:uai}) we have
\[p(d) = c_d \cdot \prod_{w \in V} p(w|d)^{f_d(w)} \textrm{and}\  p(w|d) = \sum_{t \in T} p(w|t) \cdot p(t|d).\footnote{The factor $c_d = (\sum_{w \in V} f_d(w))! / \prod_{w \in V, f_d(w) > 0} f_d(w)!$ accounts for the underlying ``bag of words model'' where word order is ignored. It is usually omitted in publications because if two approaches are compared, the expression turns out to be an identical factor for both approaches (\cite{buntine2006}). We therefore also set $c_d := 1$.}\]
The $n$ topic-word distributions form a corresponding topic model
$\phi = \{ \phi_t \}$.  Each $\phi_t = p(w|t)_{w \in V}$ represents the essence
of a topic, where $t$ itself is just for reference.

As a more sophisticated Bayesian approach, LDA puts all potential
topic-word distributions under a Dirichlet prior $\beta$ in order
to determine $p(D)$ (\cite{blei:2003:lda:944919.944937}). In this
case, an approximation of
\begin{equation} 
\label{eq:lda1} 
\Phi = argmax_\phi ((\prod_{t \in T} p(\phi_t)) \cdot \prod_{d \in D} p(d| \phi, \alpha \textbf{m}))\ \textrm{with}\ \phi_t \sim Dirichlet(\beta)
\end{equation}
may be considered a topic model (\cite{blei:2003:lda:944919.944937}). Hereby, $\alpha \textbf{m}$ is an additional Dirichlet prior to determine 
\begin{equation} 
\label{eq:lda2} 
p(d|\phi, \alpha \textbf{m}) = \int p(\theta_d) \cdot \prod_{w \in V} (\sum_{t \in T} \phi_t(w) \cdot \theta_d(t))^{f_d(w)} d\theta_d\ 
\textrm{with}\ \theta_d \sim Dirichlet(\alpha \textbf{m}).
\end{equation}
Alternatively to the $argmax$ operator, $\phi$ may be integrated out
leading to a corresponding point estimate for $\Phi$
(\cite{griffiths-steyvers:2004:pnas}).

Considering training results, $\Phi$ plays the same role as a
distribution $p(V|t)$ under pLSI.  With this in mind, \emph{we often
  use the letter $\Phi$ for topic models regardless of the underlying
  modeling approach.} A similar concession holds for document-topic
distributions $p(T|d)$.

There exist several methods and various derived algorithms
readily available to approximate $\Phi$ under LDA including
variational Bayes, MAP estimation and Gibbs sampling (e.g.~see
\cite{Asuncion:2009:SIT:1795114.1795118}).



\subsubsection{Hyper Parameter Optimization for LDA}
\label{ldahparam}

LDA is a very successful method, but suffers from the need for setting
several hyper parameters.
This sections gives a brief overview of the issue as relevant for
evaluations in Section \ref{sec:evaluation}.

Besides the number of topics $n$, standard LDA has two hyper
parameters that must be adjusted for model computation
(\cite{blei:2003:lda:944919.944937,conf/nips/wallachmm09,Asuncion:2009:SIT:1795114.1795118}):
\begin{itemize}
\item The vector $\alpha \textbf{m} \in \Re^n$ with $\sum_{i = 1}^n
  \textbf{m}_i = 1$, $\textbf{m}_i > 0$ for all $i$ and
  $\alpha > 0$ where $\alpha$ is called the \emph{concentration parameter}:
  $\alpha \textbf{m}$ parametrizes which
  document-topic distributions $\theta_d$ from Equation \ref{eq:lda2} are more or
  less probable (regardless of $d$). For practical concerns $\alpha \textbf{m}$ is often
  set with $\textbf{m}_i = 1/n$. This case is called
  ``symmetric'' since the concentration parameter $\alpha$
  remains as the only degree of freedom for $\alpha \textbf{m}$.
\item The vector $\beta\in \Re^{|V|}, \beta_{i} > 0$ for all $i$  (sometimes
  also named $\eta$): $\beta$ parametrizes, which topic-word distributions $\phi_t$ from Equation
  \ref{eq:lda1} are more or less probable (regardless of $t$). $\beta$ is usually kept symmetric.
\end{itemize}
When applying LDA, there are different approaches to determine
reasonable values for $n$, $\alpha$ and $\beta$: $n$ is often varied
via a parameter search (e.g. in \cite{blei:2003:lda:944919.944937,
  griffiths-steyvers:2004:pnas, Asuncion:2009:SIT:1795114.1795118}) with a
range typically between 10 and 1000 and a step size of 10. The
optimization criterion is high log probability or equivalently low perplexity for
held out test documents $D_{test}$.

Regarding the optimization of $\alpha \textbf{m}$, the following options are of practical relevance:
\begin{itemize}
\item If one decides for a symmetrical $\alpha$, a hyper parameter
  search may be performed (\cite{Asuncion:2009:SIT:1795114.1795118}).
The optimization goal is the same as for $n$, but one does not use
test documents as $\alpha$ is considered a more integral part of the
training process.

\item A simpler approach for a symmetrical $\alpha$, well established
  in practice, is to apply a \emph{heuristic} from
  \cite{griffiths-steyvers:2004:pnas} by setting $\alpha = 50/n$.

\item Another technique is to (re-)estimate $\alpha \textbf{m}$ as
  part of an EM procedure. E.g. in
  \cite{Asuncion:2009:SIT:1795114.1795118}, the (re-)estimation of
  $\alpha \textbf{m}$ is based on an initially computed topic model
  $\Phi_1$. The updated $\alpha \textbf{m}$ can in turn be used to
  compute an updated model $\Phi_2$ (while using $\Phi_1$ as a
  starting point to compute $\Phi_2$) an so forth. After several
  iterations of such alternating steps, the models $\Phi_i$ as well as
  $\alpha \textbf{m}$ converge. \cite{minka:2000:tr} provides a
  theoretical basis for the estimation of Dirichlet parameters via
  sample distribution data. In case of $\alpha \textbf{m}$, these are
  (samples of) estimated distributions $p(T|d)$ as computed along with
  an intermediate model $\Phi_i$. To do so,
  \cite{Asuncion:2009:SIT:1795114.1795118} leverage Equation 55 from
  \cite{minka:2000:tr} in the EM procedure and coined for this particular
  E-step the name ``Minka's update''. Minka's update can be implemented under a
  symmetrical as well as under an asymmetric $\alpha$.
\end{itemize}

Concerning $\beta$, there exists similar alternatives as for $\alpha
\textbf{m}$. A related heuristic for a symmetrical $\beta$ from
\cite{griffiths-steyvers:2004:pnas} is $\beta =
0.1$. \cite{conf/nips/wallachmm09} report that an \emph{asymmetric
  $\beta$ optimization offers worse predictive performance than its
  symmetrical counter part}, but they also stress \emph{the importance
  of the asymmetric $\alpha$ case} for topic model quality.

Later, when comparing LDA against Topic Grouper in
Section~\ref{sec:evaluation}, we refer to the heuristics for $\alpha$
and $\beta$ from \cite{griffiths-steyvers:2004:pnas} as ``\emph{LDA with
  Heuristics}''. We also include an optimization for an asymmetric $\alpha\textbf{m}$
combined with a symmetric $\beta$ optimization using Minka's update
and call it ``\emph{LDA Optimized}''.  We include both approaches in
our evaluation as extreme variants for LDA hyper parametrization: the
former one being straight forward and efficient; the latter one
offering higher predictive performance but also incurring substantial
computational overhead due to intertwined approximation procedures.
We use Gibbs sampling according to \cite{griffiths-steyvers:2004:pnas} in
order to compute intermediate topic models $\Phi_i$ as described above and
a final model, respectively  $\Phi$.\footnote{Although intricate, details on hyper parameter settings matter: Some publications compare approaches to LDA but for example, leave it unclear whether  $\alpha$ is kept symmetric or if it is optimized. 
E.g., \cite{tan-ou:2010:iscslp} report that ``basic  LDA  fails'' to successfully learn a solution for the kind of data we use in Section 
\ref{syn_data}. In comparison, we found that LDA succeeds in this case if its hyper parameters are set accordingly.}
 
\subsubsection{Hierarchical Topic Models}

Traditional topic models create flat topics; however, it may be more
appropriate to have a hierarchy comprising multiple levels of
super-topics and increasingly specialized sub-topics. To address this,
topic model extensions based on trees and directed acyclic graphs have
been proposed. 

One of the early attempts towards hierarchical topic
models is \cite{hofmann:1999:ijcai}'s Cluster Abstraction Model (CAM),
using an instance EM with annealing:
Leaf nodes of a hierarchy are generated first via probabilistic soft clustering of documents.
Inner nodes form latent sources of each word occurrence in a document such that a respective inner node is
the ancestor of a leaf cluster in which the document is placed. The latent sources are subject to probabilistic modeling based on the hierarchy's leaves. Experiments indicate that top probability words in inner nodes form topical abstractions
of the document clusters they subsume.

\cite{segal-koller-ormoneit:2002:nips}'s \emph{Probabilistic
  Abstraction Hierarchies} (PAH) is another model based on the EM algorithm: it jointly optimizes cluster
assignment, class-specific probabilistic models (CPMs) which are
taxonomy nodes and the taxonomy structure. The latter two are globally
optimized. The authors state that ``data is generated only at the leaves of the tree, so that a model
basically defines a mixture distribution whose components are the CPMs at the leaves of the tree.''
They offer a brief evaluation including a predictive performance comparison of PAH with hierarchical clustering on gene expression data.

\cite{blei-etal:2003:nips} discuss an extension of the ``Chinese
restaurant process'' (CRP) from \cite{me22}: Their so-called ``\emph{nested Chinese restaurant process}'' (nCRP) allows for inferring hierarchical mixture models while permitting uncertainty about branching factors.
Based on the nCRP, the authors propose \emph{Hierarchical LDA} (hLDA) to estimate topic trees of a given depth $L$. Documents are thought to be generated by first choosing a path of length $L$ along a tree and then mixing the document's topics via the chosen path where each path node represents a topic to be inferred.
The corresponding document-topic distribution is subject to a Dirichlet distribution with prior $\alpha$. Under hLDA, higher level topics tend to be common across many documents, but do not necessarily form semantic generalizations of lower level topics. I.e., the model tends to push stop words and function towards the root of tree and rather domain-specific words towards the leaves.
Besides $L$ and $\alpha$, hLDA requires a prior $\gamma$ affecting the branching factor of estimated trees and a prior $\eta$, which is equivalent to $\beta$ under LDA.

The \emph{Hierarchical Dirichlet Process} (HDP) by
\cite{teh-etal:2006:jamstatassoc} is a framework for two or more layered Dirichlet processes (DPs), where
a first-level DP produces the parameters for $J$ second level DPs which in turn create mixture components to explain $J$ groups of data. A merit of the HDP is that the number of mixture components on the second level must not be set while still enabling a degree of sharing of mixture components between the groups.
E.g. with regard to topic modeling, the authors apply the HDP in order to infer the number of \emph{flat} topics on a small-sized document collection along with a respective topic model. 
The HDP still mandates hyper parameters similar to $\alpha$ and $\beta$ under LDA.
\cite{wang-paisley-blei:2011:jmlr} present a faster inference algorithm for HDP, which scales up to larger dataset sizes.

The \emph{Pachinko Allocation Model} (PAM)
from \cite{wei-mccallum:2006:icml} is a hierarchical
topic model based on multiple Dirichlet processes. The PAM requires a directed acyclic graph (DAG) as a prior,
where leaf nodes correspond to words from the vocabulary, parents of leaf nodes correspond to \emph{flat}, word-based topics and other nodes represent 
mixture components over their children's mixture components.
A topic for a word occurrence of a document is sampled by considering all paths from the root to the leaves' parents. 
Moreover, the mixture components of all inner nodes are subject to Dirichlet distributions.
Due to this structure, higher level nodes in the graph form abstractions of topic mixtures across documents and therefore
capture topic correlations. As respective super-topics represent mixes over topics, the authors do not offer a labeling scheme for them.
Besides the basic graph structure, the PAM has similar hyper parameters as LDA including $\alpha$, $\beta$ and the number of word-based topics $n$. Furthermore, $\alpha$ forms of a set vectors, one for each inner node, which are estimated as part of PAM's inference process.

The \emph{recursive Chinese Restaurant Process} (rCRP) from \cite{Kim:2012:MTH:2396761.2396861} is another extension of the CRP to infer hierarchical topic structures. In contrast to hLDA, the sampling of a document-topic distribution is generalized in a way that permits a document's topics to be drawn from the entire (hierarchical) topic tree, not just from a single path. Regarding document-topic assignments, the rCRP makes the drawing of topics deeper in the tree more unlikely and estimates the branching factor of a topic tree node similarly to a regular CRP. The topic-word distributions of a tree-node are controlled via a Dirichlet with a symmetrical prior $\beta^k$, where $\beta  < 1$ and $k$ is the depth of the node. As the prior gets smaller with increasing depth, the resulting distributions get more peaked, which facilitates the production of more specific topic towards the leaves. A CRP based on a scalar prior $\alpha$ controls how words from a document are assigned to topics and another scalar prior $\gamma$ controls the inferred depth and branching factor of the topic tree under the rCRP. An experimental analysis and examples of inferred topics indicate that the approach alleviates well-known drawbacks of hLDA including the one mentioned above.

The \emph{Nested Hierarchical Dirichlet Process} (nHDP) from \cite{6802355} is perhaps the most sophisticated approach to produce tree-structured topics on the basis of DPs: 
Based on \cite{blei-etal:2003:nips} it uses the nCRP to produce a global topic tree. Every document obtains its specific topic tree which is derived from the global tree via an HDP. Hence, the HDP ensures a degree of sharing of topics between documents and allocates document-level topics based on DPs associated with the nodes of the global topic tree. To sample of a word's topic from the document-level topic tree the nHDP descends through that tree and may stop at any node. Stopping or progressing is a random event based on node-related probabilities drawn from a beta distribution with hyper parameters $\gamma_1$ and $\gamma_2$. The approach also mandates a hyper parameter $\alpha$ for its basic nCRP and $\beta$ for document-level trees. The authors provide efficient inference procedures and offer impressive results on small as well as very large text datasets, where the vocabulary on the large datasets is reduced to about 8,000 words.

An apparent commonality of the presented approaches is the need for hyper parameters---usually several scalars.
This also holds for the \emph{Hierarchical Latent Tree Analysis} (HLTA) from \cite{liu-zhang-chen:2014:ecml} 
and the \emph{Hierarchical PAM} (HPAM) from \cite{mimno-li-mccallum:2007:icml}.
An analyst applying a related approach may therefore struggle with its complexity and with setting the hyper parameters. 
Although some of the above-mentioned solutions scale up to large datasets, the resulting topic trees remain rather shallow.
In contrast, Topic Grouper offers deep trees and requires no hyper parameters. Deeper tree nodes cover only small sets of words and 
tend to become more specific. The fact that word sets are disjunctive at every tree level may ease
topic interpretation but it also imposes a limitation with regard to polysemic words. Related pros and cons will be addressed further in Sections \ref{viz} and \ref{sec:discussion}.

\subsection{Evaluation Regimes} 
\label{perplexity}

Since typically, there exists no ground truth regarding topic models, a well-established \emph{intrinsic}
evaluation scheme is to compute the
log probability for test documents $d \in D_{test}$ withheld from the training data. In this context, estimating (the logarithm of) $p(d|\Phi, \alpha \textbf{m})$ via an LDA topic model $\Phi$ with its Dirichlet prior $\alpha \textbf{m}$ 
is a non-trivial problem in itself.
We follow \cite{wallach-etal:2009:icml}, who determine this quantity conceptually as follows: 
\begin{equation} 
\label{eq:ldadoc2est} 
p(d|\Phi, \alpha \textbf{m}) = \int p(\theta_d) \cdot \prod_{w \in V} (\sum_{t \in T} \Phi_t(w) \cdot \theta_d(t))^{f_d(w)} d\theta_d\ \textrm{with}\ \theta_d \sim Dirichlet(\alpha \textbf{m})
\end{equation}
Note that apart from using $\Phi$ instead of $\phi$ from Section~\ref{sec:concepts}, Equation \ref{eq:ldadoc2est} and Equation \ref{eq:lda2} are the same.

\cite{wallach-etal:2009:icml} also examine different approximation
methods for Equation \ref{eq:ldadoc2est} and introduce their so-called
``left-to-right'' method. \cite{buntine2009} presents a refined and
unbiased version of ``left-to-right'' named ``left-to-right
sequential''. Regarding LDA, we report results based on the latter
algorithm since it acts as a gold standard estimation for Equation
\ref{eq:ldadoc2est} (see~\cite{buntine2009}).

Like \cite{blei:2003:lda:944919.944937} and others we use \textit{perplexity} as 
a derived measure to aggregate the predictive power of $\Phi$ over  $D_{test}$:
\begin{equation}
\label{eq:perplexity} 
perplexity(D_{test}) := \exp (- \sum_{d \in D_{test}} \log p(d|\Phi, \alpha \textbf{m}) / \sum_{d \in D_{test}} |d|).
\end{equation}
In doing so, only words from the training vocabulary $V$ are considered, such that
the size of a test document is $|d| = \sum_{w \in V} f_d(w)$.

An intrinsic evaluation alone does not guarantee that learned topics coincide with human intuition and interpretability. 
This is particularly important when topics are consumed by humans directly rather than being utilized as an intermediate step of a machine learning or natural language processing pipeline.
\emph{Extrinsic} evaluations therefore resort to external resources to assess topic quality: for instance, \cite{DBLP:conf/nips/ChangBGWB09} describe two human experiments,
one study on \emph{word intrusion} and another one on \emph{topic intrusion}, respectively. 
In the word intrusion task subjects are asked to identify which
spurious (``intruder'') word was added to a topic in hindsight to
``pollute'' it. If subjects identify the intruder artificially
injected by the experimenter, this is a sign that the other words
making up the topic are of good quality.  
In the topic intrusion task,
subjects are asked to identify a ``rogue topic'' that has been added
to a document (i.e., topics that are not actually covered in a document).
Regarding their setting the authors find that ``surprisingly, topic
models which perform better on held-out likelihood may infer less
semantically meaningful topics''.

\cite{Newman:2010:AET:1857999.1858011} experiment with word
co-occurrence measures obtained via word statistics from WordNet,
Wikipedia and the Google search engine. They combine related values as
obtained from each pair of a topic's top words in order to compute
\emph{topic coherence}, which they define as ``average semantic
relatedness between a topic's words''.  Several variants of resulting
quality measures matched the expectation of human annotators on
respective text collections, including pointwise mutual information
(PMI).  Other than in the intruder scenario, annotators had to rate
the coherence of topics as obtained from the training phase.  Building
on this work, \cite{W13-0102} compare four
similarity functions for the automatic evaluation of topic coherence,
including the cosine similarity, Dice coefficient
and Jaccard coefficient. While Newman \textit{et al.} use PMI to
measure similarity between a topic's top words directly,
Aletras and Stevenson first map each word of a topic to a vector of co-occurring words as computed via word statistics from Wikipedia.
Afterwards, the similarity measures are applied to such word vectors in order to estimate a topic's coherence. An evaluation based on three document collections and envolving human judges shows that their approach performs better than using PMI directly.


\cite{lau-newman-baldwin:2014:eacl} build on the work from \cite{DBLP:conf/nips/ChangBGWB09} and \cite{Newman:2010:AET:1857999.1858011} and also offer a good review of extrinsic evaluations for topic models.
They use machine learning to automate the detection of intruder words and to automatically assess the degree of coherence of a topic, respectively.
While they solved the latter task successfully, the former task posed problems. Maybe surprisingly, they find that ``the correlation between the human
ratings of intruder words and observed coherence is only modest'' and give a plausible example-based explanation in their paper.

This look at extrinsic evaluation methods indicates that they are manifold and that related research is still ongoing.
We therefore rely on hold-out performance for now as a well-established and more standardized criterion.
Numerous topic modeling contributions suggest that at least a reasonable hold-out performance is a necessary criterion also for semantically meaningful topic models.
Evidence usually comes from reporting such performance results in conjunction with example topics as learned from a text collection covering general knowledge (e.g.~ see \cite{mimno-li-mccallum:2007:icml}, \cite{Kim:2012:MTH:2396761.2396861} or \cite{6802355}).
We follow this scheme, but also leverage some simple \emph{synthetic datasets} in order to examine whether a modeling approach is able to recover \emph{the true topics} governing that dataset.

\section{Topic Grouper}
\label{sec:theory}
\subsection{Model}
\label{basics}
Let $T(n) = \{t\ |\ t \subseteq V\} $ be a (topical) partitioning of $V$ such that $s \cap t = \emptyset$ for any $s, t \in T(n)$, $\bigcup_{t \in T(n)} t = V$ and $|T(n)| = n$.
Further, let the \emph{topic-word assignment} $t(w)$ be the topic of a word $w$ such that $w \in t \in T(n)$.
Note that in the following, we also make use of the variables $D$, $V$, $f_d(w)$ and $\Phi$ as specified in Section \ref{sec:concepts}.

Our principal goal is to find an \emph{optimal} partitioning $T(n)$ for each $n$ along
\[ argmax_{T(n)} q(T(n)), \textrm{with}\]
\[ q(T(n)) := \prod_{d \in D} \prod_{w \in V, f_d(w) > 0} \left( p(w | t(w)) \cdot p(t(w) | d) \right) ^{f_d(w)}.\]
The idea is that each document $d \in D$ is considered to be \emph{generated} via a simple stochastic process where a word $w$ in $d$
occurs by
\begin{itemize}
\item first sampling a topic $t$ according to a probability distribution $p(t | d)_{t \in T(n)}$,
\item then sampling a word from $t$ according to the topic-word distribution $p(w | t)_{w \in V}$
\end{itemize}
and so, the total probability of generating $D$ is proportional to $q(T(n))$.

The optimal partitioning consists of $n$ pairwise disjunctive subsets of
$V$, whereby each subset is meant to represent a topic.  By definition
every word $w$ must be in exactly one of those sets. This may help to
keep topics more interpretable for humans because they do not overlap
with regard to their words. On the other hand, polysemic words can
only support one topic, even though it would be justified to keep them
in several topics due to multiple contextual meanings. Note that the
approach considers a solution for every possible number of topics $n$
ranging between $|V|$ and one.

To further detail our approach, we set
\begin{itemize}
\item $f(w) := \sum_{d \in D} f_d(w) > 0$, since otherwise $w$ would not be in the vocabulary,
\item $|d| := \sum_{w \in V} f_d(w) > 0$, since otherwise the document would be empty,
\item $f_d(t) := \sum_{w \in t} f_d(w)$ be the topic frequency in a document $d$ and
\item $f(t) := \sum_{w \in t} f(w) = \sum_{d \in D} f_d(t)$ be the number of times $t$ is referenced in $D$ via some word $w \in t$.
\end{itemize}
Concerning $q(T(n))$ we use maximum likelihood estimations for $p(t(w) | d)$ and $p(w | t(w))$ based on $D$:
\begin{itemize}
\item $p(t(w) | d) \approx f_d(t(w)) / |d|$, which is $> 0$ if $f_d(w) > 0$,
\item $p(w | t(w)) \approx f(w) / f(t(w))$, which is always $> 0$ since $f(w) > 0$. 
\end{itemize}

Unfortunately, constructing the optimal partitionings $\{ T(n)\ |\ n = 1\dots|V| \}$ is computationally hard.
\emph{We suggest a greedy algorithm that constructs suboptimal partitionings instead}, starting with $T(|V|) := \{ \{ w \}\ |\ w \in V \}$ as step $i = 0$.
At every step $i = 1\dots|V| - 1$ the greedy algorithm joins two different topics $s,t \in T(|V| - (i - 1))$ such that
$q(T(|V| - i))$ is maximized while $T(|V| - i) = \left( T(|V| - (i - 1)) - \{s, t\} \right) \cup \{ s \cup t \}$ must hold.
Essentially, this results in an \emph{agglomerative clustering approach, where topics, not documents, form respective clusters}. 

For efficient computation we first rearrange the terms of $q(T(n))$ with a focus on topics in the outer factorization:
\[ q(T(n)) = \prod_{t \in T(n)} \prod_{d \in D, f_d(t) > 0} \left( p(t | d)^{f_d(t)} \cdot \prod_{w\in t}p(w | t)^{f_d(w)}\right) \]
The rearrangement relies on the fact that every word belongs to exactly one topic and enables the 
``change of perspective'' towards topic-oriented clustering.

We maximize $\log q(T(n))$ instead of $q(T(n))$ which is equivalent with respect to the argmax-operator. This leads to
\[ \log q(T(n)) = \sum_{t \in T(n)} \sum_{d \in D, f_d(t) > 0} (f_d(t) \cdot \log p(t | d) + \sum_{w\in t} f_d(w) \cdot \log p(w | t)) \approx \sum_{t \in T(n)} h(t) \]
with the maximum likelihood estimation
\begin{equation} 
\label{eq:hnt} 
h(t) := \sum_{d \in D, f_d(t) > 0} f_d(t) \cdot (\log f_d(t) - \log |d|) + \sum_{w\in t} f(w) \cdot \log f(w) - f(t) \cdot \log f(t).
\end{equation}
Using these formulas the best possible join of two (disjunctive) topics $s, t \in T(n)$ results in $T(n - 1)$ with
\[ \log q(T(n - 1)) \approx \log q(T(n)) + \Delta h_n, \]
\begin{equation} 
\label{eq:deltahn}
\Delta h_n := max_{s,t \in T(n)} \Delta h(s,t)\ \textrm{and}
\end{equation}
\begin{equation} 
\label{eq:deltah} 
\Delta h(s,t) := h(s \cup t) - h(s) - h(t).
\end{equation}
From the perspective of clustering procedures $-\Delta h(s,t)$ is the cluster distance between $s$ and $t$.
Note though, that it does not adhere to standard distance axioms.

\subsection{Joining Two Topics $s$ and $t$}
Considering the resulting algorithm, we can reuse $h(s)$ and $h(t)$ from prior computation steps in order to compute $h(s \cup t)$ efficiently: 
Regarding expression (\ref{eq:hnt}) from above, let $i(t) := \sum_{w\in t} f(w) \cdot \log f(w)$. 
We have 
$f_d(s \cup t) = f_d(s) + f_d(t)$, $f(s \cup t) = f(s) + f(t)$ and $i(s \cup t) = i(s) + i(t)$, and so
\begin{equation} 
\label{eq:efficienth} 
\begin{aligned}
h(s \cup t) = \sum_{d \in D, f_d(s) + f_d(t) > 0} (f_d(s) + f_d(t)) \cdot \left( \log (f_d(s) + f_d(t)) - \log |d|\right) + \\
i(s) + i(t) - (f(s) + f(t)) \cdot \log (f(s) + f(t)).
\end{aligned}
\end{equation} 
The terms $i(u)$ and $f(u)$ with $u = s, t$ will have been computed already during the prior steps of the resulting algorithm, i.e. when $t$ and $s$ were generated as topics. Thus, the computation of all sums over words $w$ can be avoided with respect to $h(s \cup t)$. This is essential for a reasonable runtime complexity.

\subsection{Initialization}
\label{initialization}
During initialization, the resulting algorithm generates all one-word topics $t \in T(|V|)$. Given $t = \{ w \}$ we have
\begin{equation} 
\label{eq:init1} 
h(\{w\}) = \sum_{d \in D, f_d(w) > 0} f_d(w) \cdot (\log f_d(w) - \log |d|).
\end{equation}
The algorithm also computes the best possible join partner $s = \{ v \}$ for some $t = \{ w \}$ and so 
\begin{equation} 
\label{eq:init2} 
\begin{aligned}
h(\{ v, w \}) = \sum_{d \in D, f_d(v) + f_d(w) > 0} (f_d(v) + f_d(w)) \cdot (\log (f_d(v) + f_d(w)) - \log |d|) + \\
i(\{ v \}) + i(\{ w \}) - (f(v) + f(w)) \cdot \log (f(v) + f(w)).
\end{aligned}
\end{equation}
The first sum in this expression is problematic because one would have to iterate over the document set to compute it. 
Using an inverted index, one can avoid looking at documents with $f_d(v) = 0$ and $f_d(w) = 0$.

\subsection{Algorithm and Complexity}
\label{complexity}
Topic Grouper can be implemented via adaptations of standard
agglomerative clustering algorithms:  Listing \ref{lst:ehac} presents
a related variant of the \emph{efficient hierarchical agglomerative
clustering} (EHAC) taken from \cite{manning:2008:iir:1394399}, which
manages a map of priority queues in order to represent evolving
clusters during the agglomeration process.  EHAC's time complexity is
in $O(k^2 \log k)$ and its space complexity in $O(k^2)$ with $k$ being
the initial number of clusters. However, this implies that the cost of
computing the distance between two clusters is in $O(1)$.  In the case of
Topic Grouper the latter cost is in $O(|D|)$ instead, because one must
compute the value of $h$ from Equation \ref{eq:efficienth}. The factor
``$\log k$'' from EHAC's original time complexity accounts for access
to priority queue elements -- in the case of Topic Grouper this is
dominated by the cost to compute $h$-values.

Putting it together, the time complexity for Listing \ref{lst:ehac} is
on the order of $|V|^2 \cdot |D|$ and its space complexity is in
$O(|V|^2)$.  In case of text, one may further assume that Heaps' Law
holds (\cite{book/heaps}): Without a fixed limit on the vocabulary, we then have about $|V|^2 \sim |D|$,
leading to a simplified time complexity estimation for Topic Grouper
roughly on the order of $|D|^2$.

The stated space complexity, $O(|V|^2)$, can be problematic if the
vocabulary is large. We devised an alternative clustering algorithm,
MEHAC, whose space complexity is in $O(|V|)$ but its \emph{expected}
time complexity is still in $O(|V|^2 \cdot |D|)$.  A drawback of MEHAC
is that in practice, it incurs a higher constant computation time
factor than EHAC. So, given sufficient memory, the EHAC variant is
preferable.  MEHAC is detailed in Appendix \ref{mehac}. 
Appendix \ref{pperformance} highlights the practical performance of both
algorithms based on example datasets.
\newpage
\lstset{numbers=left,morekeywords={new, foreach, var, foreach,
    procedure, print, insert, remove, add, null, while, true, false,
    clear},basicstyle=\tiny,escapeinside={(*}{*)}}
\begin{lstlisting}[mathescape=true,caption={Variant of Efficient Agglomerative Clustering (EHAC) for Topic Grouper\\},label=lst:ehac]
// (*\textbf{Input: $V, D, f_d(w)$ and $f(w)$ according to Section
  \ref{basics}}*) // (*\textbf{Output: Relevant changes of T -- the
  current set of topics -- printed out.}*)

// (*\textbf{Global variables}*)
var T := $\emptyset$; // Current set of topics, topics are assumed to be fully ordered (no matter how)
// Map of priority queues of topics. Each topic s from T acts as a key and maps to one queue. 
// Moreover, each queue's topics t are sorted in descending order on the basis of $\Delta h(s,t)$:
var pq[];
// Map for parameters from Equation (*\ref{eq:efficienth}*), topics from T are used as keys:
var h[], f[], i[], fd[];

// (*\textbf{Initialization step $i$ = 0}*)
foreach $w \in V$ { // Filling T
  var t := { $w$ };
  insert t into T;
  h[t] := $h(t)$ according to Equation (*\ref{eq:init1}*)
  foreach $d \in D$ { fd[(t,d)] := $f_d(w)$; }
  f[t] := $f(w)$; i[t] := $f(w) \cdot \log f(w)$;
}
print T;
foreach t $\in$ T { pq[t] := new PriorityQueue(); }
foreach s $\in$ T { // Computing initial join partners  
  foreach t $\in$ T with t > s {
    var u := s$\cup$t;
    h[u] := $h(u)$ according to Equation (*\ref{eq:init2}*);
    var $\Delta$h := h[u] - h[s] - h[t];
    add t to pq[s] on the basis of $\Delta$h;
    add s to pq[t] on the basis of $\Delta$h;
  }
}
// (*\textbf{Steps i $>$ 0 to join topics}*)
while (|T| > 1) {
  var s := $argmax_{\textrm{r} \in \textrm{T}}$ pq[r].peek.$\Delta$h; // Determine queue pq[s] with best head on the basis of $\Delta h$.
  var t := pq[s].pull; // Remove head from pq[s] and return it.
  var u := s$\cup$t;
  remove s from T; remove t from T;
  insert u into T;
  print T;
  // Update data structures:
  foreach $d \in D$ { fd[(u, d)] := fd[(s,d)] + fd[(t,d)]; clear fd[(s,d)], fd[(t,d)]; }
  f[u] := f[s] + f[t]; i[u] := i[s] + i[t];	
  clear pq[s], pq[t], h[s], h[t], f[s], f[t], i[s], i[t];
  foreach v $\in$ T { remove s from pq[v]; remove t from pq[v]; }
  // Update join partners for u:
  foreach r $\in$ T with r $\neq$ u {
    var v = r$\cup$u;
    h[v] := $h(v)$ according to Equation (*\ref{eq:efficienth}*);
    var $\Delta$h := h[v] - h[r] - h[u];
    add v to pq[u] on the basis of $\Delta$h;
    add u to pq[v] on the basis of $\Delta$h;
  }
}
\end{lstlisting}

\section{Experiments} \label{sec:evaluation}

\subsection{Synthetic Data}
\label{syn_data}

This section provides a first evaluation of Topic Grouper using
simple synthetically generated datasets. As
the true topics $S$ are known (i.e. having gold data), this allows us to consider \emph{error rate} as
a quality measure and to examine some basic qualities of
our approach: The idea is to compare a model $\Phi$
against the \emph{true} topic-word distributions used to generate
a dataset.

The following definition of error rate $err$ assumes that the perfect
number of topics is already known, such that $|T| := |S|$ is preset
for training. The order of topics in topic models is unspecified, so
we try every \emph{bijective} mapping $\pi : T \rightarrow S$ when
comparing a computed model $\Phi$ with a true model $\tilde{p}(V|S)$
and favor the mapping that minimizes the error:
\[ err := \min_{\pi} \frac{1}{2 |T|} \sum_{t \in T} \sum_{w \in V} |\Phi_t(w) - \tilde{p}(w| \pi(t))|.\]
The measure is designed to range between 0 and 1, where 0 is
perfect. Considering a mapping $\pi$, every topic may
contribute equally to lower the error rate. The factor $1/2$ avoids
double counting, since a quantity $\Phi_t(w)$ exceeding $\tilde{p}(w|
\pi(t))$ will be missing for other words $w'$, i.e. $\Phi_t(w')$ will
then be too low.

\subsubsection{Datasets According to \cite{tan-ou:2010:iscslp}}
\label{twcdataeval}

We use a simple synthetic data generator as introduced in
\cite{tan-ou:2010:iscslp}: It is based on $|V| = 400$ (artificial)
words equally divided into $4$ \emph{disjoint} topics $S = \{ s_1,
\ldots, s_4 \}$. The words are represented by numbers, such that
$0\ldots99$ belongs to $s_1$, $100\ldots 199$ to $s_2$ and so on.

Concerning the 100 words of a topic $s_i$, the topic-word
distribution $\tilde{p}(w|s_i)_{w \in V}$ is drawn independently for each topic
from a Dirichlet distribution with a symmetric prior $\tilde{\beta} =
1/100$, such that $\sum_{w = (i - 1) \cdot 100}^{i \cdot 100 - 1}
\tilde{p}(w|s_i) = 1$. A resulting dataset holds 6,000 documents with
each document consisting of 30 word occurrences. A document-topic
distribution $\tilde{p}(w|d)_{w \in S}$ is drawn independently for each document
via a Dirichlet with the prior $\tilde{\alpha} \tilde{\textbf{m}} =
(5, 0.5, 0.5, 0.5)^\top$, where topic 1 with $\tilde{\alpha}
\tilde{\textbf{m}_1} = 5$ is meant to represent a typical ``stop word
topic'', which is more likely than other topics.

To generate a word occurrence for a document $d$, the occurrence's
topic $s_i$ is first drawn via $\tilde{p}(S|d)$. Then, the word is
drawn via $\tilde{p}(V|s_i)$.
For the results from below, we generated two random datasets (``1''
and ``2'') this way, where each has its specific topic-word
distributions $\tilde{p}(V|s_i)$.

\subsubsection{Results}

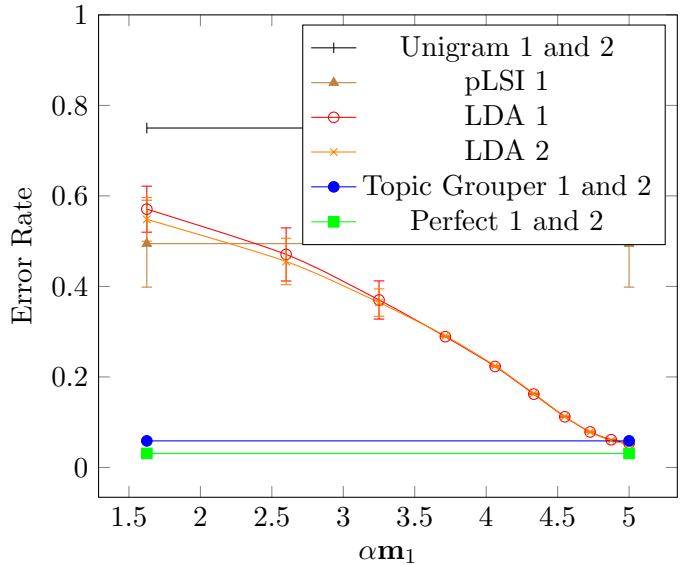
\begin{figure}
\begin{center}
\begin{tikzpicture} 
\begin{axis}[
  xlabel=$\alpha \textbf{m}_1$,
  ylabel=Error Rate,
	ymax=1,
	height=8cm
]

\addplot[smooth,color=black,mark=|] table [x=alpha1, y=errorRate, col sep=semicolon] {./csv/twc/TWCUnigramPerplexityChangeAlphaExp.csv};
\addlegendentry{Unigram 1 and 2}

\addplot[smooth,color=brown,mark=triangle*,error bars/.cd, y dir = both, y explicit] table [x=alpha1, y=errorRateAvg, y error=errorRateStdDev, col sep=semicolon] {./csv/twc/TWCPLSAPerplexityChangeAlphaExp.csv};
\addlegendentry{pLSI 1}


\addplot[smooth,color=red,mark=o,error bars/.cd, y dir = both, y explicit] table [x=alpha1, y=errorRateAvg, y error=errorRateStdDev, col sep=semicolon] {./csv/twc/TWCLDAPerplexityChangeAlphaExp.csv};
\addlegendentry{LDA 1}

\addplot[smooth,color=orange,mark=x,error bars/.cd, y dir = both, y explicit] table [x=alpha1, y=errorRateAvg, y error=errorRateStdDev, col sep=semicolon] {./csv/twc/TWCLDAPerplexityChangeAlphaExp2.csv};
\addlegendentry{LDA 2}

\addplot[smooth,color=blue,mark=*] table [x=alpha1, y=errorRate, col sep=semicolon] {./csv/twc/TWCTGPerplexityChangeAlphaExp.csv};
\addlegendentry{Topic Grouper 1 and 2}

\addplot[smooth,color=green,mark=square*] table [x=alpha1, y=errorRate, col sep=semicolon] {./csv/twc/TWCPerfectPerplexityChangeAlphaExp.csv};
\addlegendentry{Perfect 1 and 2}

\end{axis}
\end{tikzpicture}
\caption{Error Rate Depending on $\alpha \textbf{m}_1$ for Two Datasets Generated According to \cite{tan-ou:2010:iscslp}}
\label{ldasymresult}
\end{center}
\end{figure}

Figure \ref{ldasymresult} shows the error rate of LDA as well as Topic
Grouper for the two datasets from Section \ref{syn_data}.  The values
were produced via a 75\% random sub-sample or taken from each dataset
for training, respectively. The remaining 25\% were used as test data
in order to compute perplexity --- corresponding results can be found in Appendix \ref{perplexity_syn}.

Regarding LDA, the depicted values are averaged across 50 runs per
data point,
whereby the random seed for the
Gibbs sampler was changed for every run. The symmetric hyper parameter
$\beta$ was optimized using Minka's update (see Section
\ref{ldahparam}).

LDA's $\alpha \textbf{m}$ changes along the X axis such that $\alpha =
\tilde{\alpha} = 6.5$ and $\textbf{m}_2 = \textbf{m}_3 = \textbf{m}_4$
always hold. The results stress the importance of hyper parameter
choice for model quality under LDA with regard to $\alpha
\textbf{m}$. This conforms to respective findings from
\cite{conf/nips/wallachmm09}. Note that a symmetric $\alpha
\textbf{m}$ with $\alpha \textbf{m}_1 = 1.625$ fails to deliver low
error rates.  LDA performs better as the $\alpha \textbf{m}$
approaches the true $\tilde{\alpha} \tilde{\textbf{m}}$, which governs
the datasets.

\emph{In this setting, Topic Grouper delivers good error rates right away.}  As its results are independent of $\alpha
\textbf{m}$ and $\beta$ but also deterministic, they are included as a horizontal line.

We also added results for pLSI as an alternative approach introduced
by \cite{hofmann:1999:uai} (where dataset 2 is omitted for visual
clarity): pLSI attains only mediocre and volatile results, heavily
depending on its random initialization values.  We therefore excluded it
from evaluations on other datasets from below.

The unigram model simply sets $\Phi_t(w) := f(w) / \sum_{w \in V}
f(w)$ for any $t$. For completeness and for reference, we finally
added a theoretically ``perfect model'': It determines the
topic-word probabilities on the basis of the training data while
using the perfect topic-word assignment as known from data
generation.

It is worth mentioning that we ran additional experiments with many
other configurations for the data generator from Section
\ref{twcdataeval}: E.g., we varied the number of topics, words per
document, vocabulary size, number of documents and $\tilde{\alpha}
\tilde{\textbf{m}}$ but kept up the unique topic-word assignment as
part of the generation. Such obtained results were analogous to the
reported ones.

We also compared how LDA's and Topic Grouper's error rates drop with an increasing number of training documents generated according to Section \ref{twcdataeval}. In favour of LDA, we set $\alpha \textbf{m} := \tilde{\alpha} \tilde{\textbf{m}}$. With the number documents ranging between 500 and 10.000 both approaches attained about similar performance (not depicted).

%

\begin{figure}
\begin{center}
\begin{tikzpicture}
\begin{axis}[scaled ticks=false, tick label style={/pgf/number format/fixed},
  axis y line*=left,
	x dir=reverse,
  xlabel=Number of Topics $n$,
  ylabel=$\Delta h_n$,
	width=10cm,
	height=6cm,
  legend style={at={(0.5,0.5)},anchor=east}
]
\addplot[color=blue,mark=o] table [x=ntopics, y=improvement, col sep=semicolon] {./csv/twc/TWCLikelihoodTG.csv};
\label{plot_one}
\addlegendentry{}
\end{axis}
\begin{axis}[scaled ticks=false, tick label style={/pgf/number format/fixed},
  axis y line*=right,
  axis x line=none,
	x dir=reverse,
  ylabel=$\Delta h_n / \Delta h_{n+1}$,
	width=10cm,
	height=6cm,
  legend style={at={(0.5,0.5)},anchor=east}
]
\addlegendimage{/pgfplots/refstyle=plot_one}\addlegendentry{$\Delta h_n$}
\addplot[color=red,mark=x] table [x=ntopics, y=improvementratio, col sep=semicolon] {./csv/twc/TWCLikelihoodTG.csv};
\addlegendentry{$\Delta h_n / \Delta h_{n-1}$}
\end{axis}
\end{tikzpicture}
\caption{$\Delta h_n$ from Equation \ref{eq:deltahn} Depending on the Number of Topics $n$ for a Dataset Generated According to \cite{tan-ou:2010:iscslp}}
\label{likelihoodresult}
\end{center}
\end{figure}
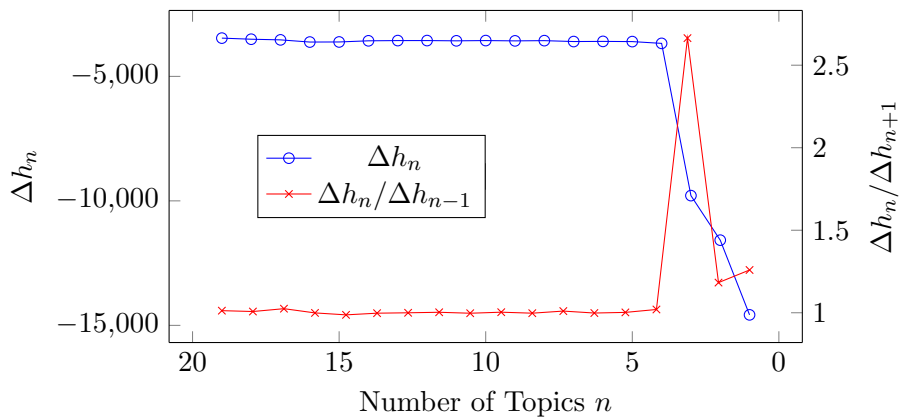

Figure \ref{likelihoodresult} illustrates how $\Delta h_n$ from
Equation \ref{eq:deltahn} can be used as a suitable measure to
determine a good number of topics in the context of Topic Grouper.
Here, the sudden drop of $\Delta h_n$ at $n = 3$ means that at least
four topics are required to model the data accordingly. A similar
approach is often taken for LDA: E.g. \cite{griffiths-steyvers:2004:pnas}
visualize the log probability $\log p(D)$ for the training dataset $D$ under LDA where the number of topics $n$ is varied.
While under LDA a separate training run is required for every
$n$, Topic Grouper assesses all potential values of $n$ between $|V|$
and 1 within a single run.

\subsection{Real-World Datasets}
\label{realworlddata}

This section reports perplexity results for two retail datasets and
two text-based datasets. The log probability for test documents is estimated as described in Section \ref{perplexity} and perplexity is computed via Equation \ref{eq:perplexity}.

\emph{Fortunately this approach can be applied to models computed via LDA and Topic Grouper alike}: In the latter case, we set $\Phi_t(w) :=
\delta_{t(w), t} \cdot f(w) / f(t)$ with $t(w)$ being the topic to
which $w$ belongs and $f(w) / f(t(w))$ being the maximum likelihood
estimate.\footnote{$\delta$ is the Kronecker symbol.} (This implies
that $\Phi_t(w) = 0$ if $t(w) \neq t$.) As there is no predefined
prior $\alpha \textbf{m}$ under Topic Grouper, we simply set
$\textbf{m}_t = f(t) / \sum_t f(t)$ -- the maximum likelihood estimate
for $p(t)$. Finally, we determine a suitable value for the
concentration parameter $\alpha$ via an interval search with the
optimization goal being low perplexity on the training data. The such
obtained $\alpha \textbf{m}$ is used during test.
We believe the approach is fair because \emph{it focuses on the
  quality of a topic model} $\Phi$ regardless of its underlying
training method. Due to $\theta$'s Dirichlet, it also avoids
(additional) smoothing schemes for Topic Grouper.

\subsubsection{Retailing}

Regarding retailing, a shopping basket or an order are
equivalent to a document.  Articles correspond to words from a
vocabulary and item quantities transfer to word occurrence frequencies
in documents. In this context, topics represent groups of articles as
typically bought or ordered together. Therefore, inferred
topic models may be leveraged to optimize sales-driven catalog
structures, to develop layouts of product assortments
(\cite{CHEN2007976}) or to build recommmender systems
(\cite{Wang:2011:CTM:2020408.2020480}).

The ``Online Retail'' dataset is a ``transnational dataset which
contains all the transactions occurring between 01/12/2010 and
09/12/2011 for a UK-based ... online retail'' obtained from the UCI
Machine Learning Repository (\cite{Chen2012}).\footnote{See
  \url{https://archive.ics.uci.edu/ml/datasets/Online+Retail}.} We
performed data cleaning by removing erroneous and inconsistent orders.
Item quantities are highly skewed with about 5\% above 25, some
reaching values of over 1,000. This is due to a mixed customer base
including consumers and wholesalers. We therefore excluded all order
items with quantities above 25 to focus on small scale (parts of)
orders.  We randomly split such preprocessed orders into 90\% training
and 10\% test data, keeping only articles that were ordered at least
10 times in the training data. The resulting training dataset covers
$|V| = 3,464$ articles, $|D| = 17,086$ orders and 427,150 order
items. The resulting average sum of item quantities per order is about
154.

\begin{figure}
\begin{center}
\begin{tikzpicture} 
\begin{axis}[
  xlabel=Number of Topics,
  ylabel=Perplexity,
	width=10cm,
	height=7cm
]
\addplot[smooth, color=red, mark=o] table [x=topics, y=perplexityLR, col sep=semicolon] {./csv/perplexity/OnlineRetail/OnlineRetailLDAPerplexityExperiment.csv};
\addlegendentry{LDA with Heuristics}

\addplot[smooth, color=green, mark=x] table [x=topics, y=perplexityLR, col sep=semicolon] {./csv/perplexity/OnlineRetail/OnlineRetailLDAPerplexityExperimentOpt.csv};
\addlegendentry{LDA Optimized}

\addplot[smooth,color=blue, mark=*] table [x=topics, y=perplexity, col sep=semicolon] {./csv/perplexity/OnlineRetail/OnlineRetailTGLRPerplexityExperiment.csv};
\addlegendentry{Topic Grouper}
\end{axis}
\end{tikzpicture} 
\caption{Perplexity on the Preprocessed Online Retail Dataset}
\label{onlineretailresult}
\end{center}
\end{figure}
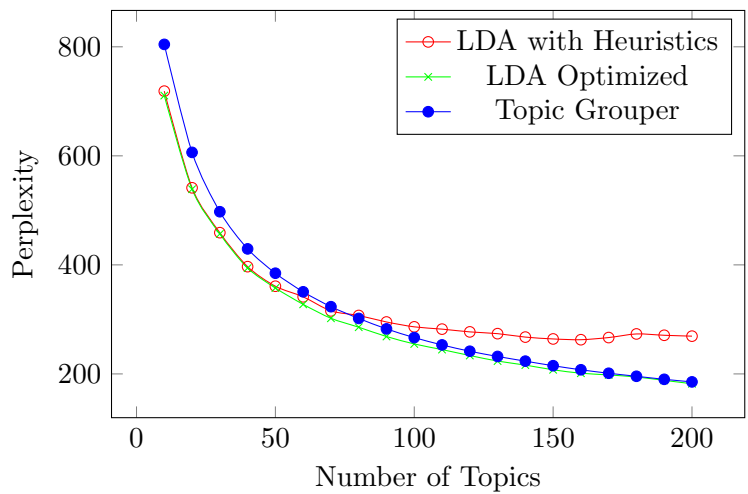

Figure \ref{onlineretailresult} shows that optimized LDA and Topic
Grouper are closely matched beyond 80 topics with optimized LDA
performing slightly better. In comparison, the performance of LDA with
heuristics begins to degrade at 80 topics. Topic Grouper is
competitive although its underlying topic model is more restrained (as
each article or word belongs to exactly one topic, respectively).

The ``Ta Feng'' dataset was published on the ACM RecSys
Wiki\footnote{See \url{http://www.recsyswiki.com}.}: It captures
shopping baskets of consumers from a Taiwanese grocery store collected
over four months between 2000 and 2001. It covers 23,812 articles and
119,578 shopping baskets but the average number of goods in a basket
is only about 9.5 with about 6.8 different articles. For data
cleaning, we removed unlikely item quantities above 50 from shopping
baskets.
Again we split the left-over data based on a 90\% to 10\% ratio, keeping only articles that were bought at least 20 times in the training data. This way we ended up with $|V| = 7,893$ articles left for training.

\begin{figure}
\begin{center}
\begin{tikzpicture} 
\begin{axis}[
  xlabel=Number of Topics,
  ylabel=Perplexity,
	width=10cm,
	height=7cm
]
\addplot[smooth, color=red, mark=o] table [x=topics, y=perplexityLR, col sep=semicolon] {./csv/perplexity/TaFeng/TaFengLDAPerplexityExperiment.csv};
\addlegendentry{LDA with Heuristics}

\addplot[smooth, color=green, mark=x] table [x=topics, y=perplexityLR, col sep=semicolon] {./csv/perplexity/TaFeng/TaFengLDAPerplexityExperimentOpt.csv};
\addlegendentry{LDA Optimized}

\addplot[smooth,color=blue, mark=*] table [x=topics, y=perplexity, col sep=semicolon] {./csv/perplexity/TaFeng/TaFengTGLRPerplexityExperiment.csv};
\addlegendentry{Topic Grouper}
\end{axis}
\end{tikzpicture} 
\caption{Perplexity on the Preprocessed Ta Feng Dataset}
\label{tafengresult}
\end{center}
\end{figure}
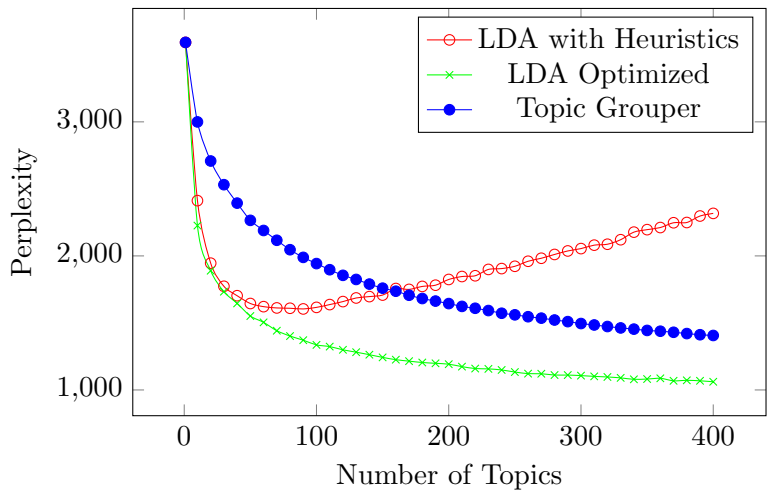

Figure \ref{tafengresult} shows the respective perplexity results. LDA
Optimized clearly dominates but Topic Grouper surpasses LDA with
Heuristics at about 180 topics. LDA with Heuristics fails at higher
topic numbers due to inappropriate hyper parameter setting.

\subsubsection{Text}

The first of two examples is a subset of the TREC AP corpus containing
20,000 newswire articles.\footnote{See
  \url{https://catalog.ldc.upenn.edu/LDC93T3A}.} We performed (Porter)
stemming and kept every stem that occurs at least five times in the
dataset. Moreover, we removed all tokens containing non-alphabetical
characters or being shorter than three characters. Again we split the
left-over documents on a 90\% to 10\% basis and only kept words
occurring at least five times in the training data. This led to $|V| =
25,047$ words left for training.

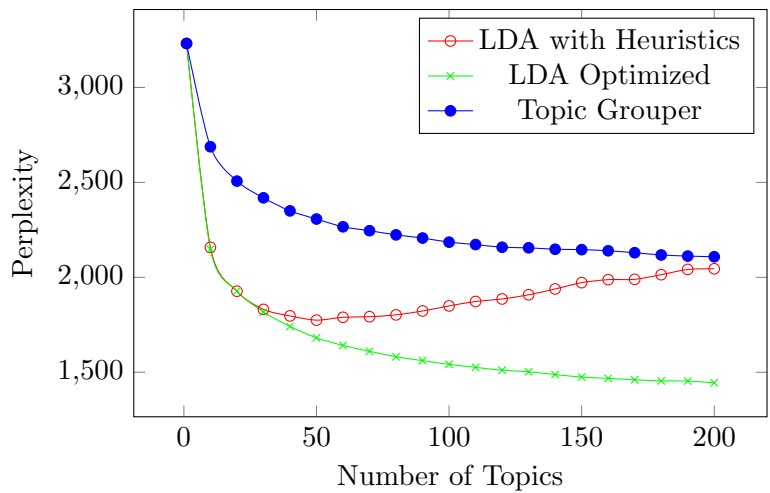
\begin{figure}
\begin{center}
\begin{tikzpicture} 
\begin{axis}[
  xlabel=Number of Topics,
  ylabel=Perplexity,
	width=10cm,
	height=7cm
]
\addplot[smooth, color=red, mark=o] table [x=topics, y=perplexityLR, col sep=semicolon] {./csv/perplexity/APLarge/APLDAPerplexityExperiment.csv};
\addlegendentry{LDA with Heuristics}

\addplot[smooth, color=green, mark=x] table [x=topics, y=perplexityLR, col sep=semicolon] {./csv/perplexity/APLarge/APLDAPerplexityExperimentOpt.csv};
\addlegendentry{LDA Optimized}

\addplot[smooth,color=blue, mark=*] table [x=topics, y=perplexity, col sep=semicolon] {./csv/perplexity/APLarge/APTGPerplexityExperiment.csv};
\addlegendentry{Topic Grouper}
\end{axis}
\end{tikzpicture} 
\caption{Perplexity on the Preprocessed AP Dataset}
\label{aplresult}
\end{center}
\end{figure}

Figure \ref{aplresult} shows related results: Here, Topic Grouper
performs generally worse than LDA but attains similar performance as
LDA with Heuristics beyond about 200 topics.  Despite these
differences we found that related topics generated by Topic Grouper
are reasonably conclusive and coherent. We will elaborate on this with
regard to the AP Corpus in Section \ref{viz}.

The NIPS dataset is a collection of 1,500 research publications from
the Neural Information Processing Systems Conference.  We used a
preprocessed version as is of the dataset from the UCI Machine
Learning Repository.\footnote{See
  \url{https://archive.ics.uci.edu/ml/datasets/Bag+of+Words}.} It was
already tokenized and had stop words removed but no stemming was
performed. We split the document set on a 90\% to 10\% basis and only
kept words occurring at least five times in the training data. This
way we ended up with $|V| = 8,801$ words left for training.

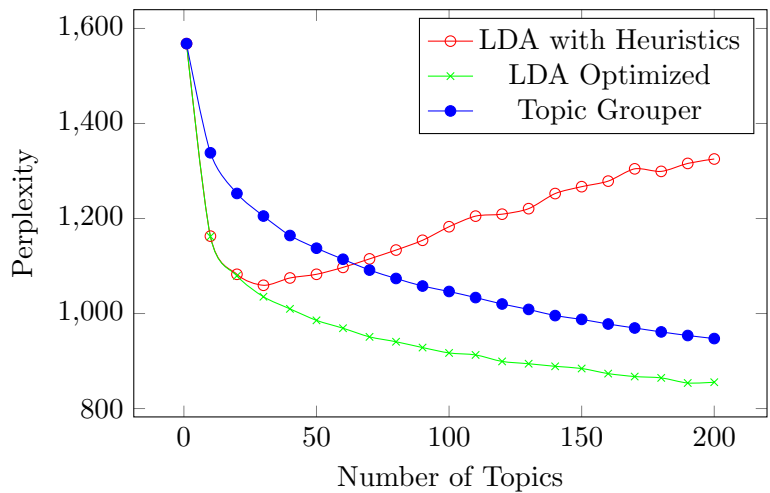
\begin{figure}
\begin{center}
\begin{tikzpicture} 
\begin{axis}[
  xlabel=Number of Topics,
  ylabel=Perplexity,
	width=10cm,
	height=7cm
]
\addplot[smooth, color=red, mark=o] table [x=topics, y=perplexityLR, col sep=semicolon] {./csv/perplexity/NIPS/NIPSLDAPerplexityExperiment.csv};
\addlegendentry{LDA with Heuristics}

\addplot[smooth, color=green, mark=x] table [x=topics, y=perplexityLR, col sep=semicolon] {./csv/perplexity/NIPS/NIPSLDAPerplexityExperimentOpt.csv};
\addlegendentry{LDA Optimized}

\addplot[smooth,color=blue, mark=*] table [x=topics, y=perplexity, col sep=semicolon] {./csv/perplexity/NIPS/NIPSTGLRPerplexityExperiment.csv};
\addlegendentry{Topic Grouper}
\end{axis}
\end{tikzpicture} 
\caption{Perplexity on the Preprocessed NIPS Dataset}
\label{nipsresult}
\end{center}
\end{figure}

Figure \ref{nipsresult} shows that LDA Optimized performs best but
Topic Grouper outperforms LDA with Heuristics beyond about 70 topics.

Together, the results of the four datasets suggest that Topic Grouper should be
considered as an option especially when words incur little ambiguity.
E.g., this tends to be the case for the retail examples, where
words represent articles without an aspect of polysemy. Also, Topic
Grouper tends to outperform LDA with Heuristics at a larger number
of topics. 

\subsection{Feature Reduction and Document Classification}
\label{featurered}

This section compares the abilities of LDA, Topic Grouper, \emph{Information
Gain} (IG) and \emph{Document Frequency} (DF) regarding feature reduction for
text classification.  In the first two cases, the idea is to exchange
word occurrences for topic occurrences and thus, to reduce feature
space dimensionality from the vocabulary size $|V|$ to the number of
topics $|T|$.  In contrast, IG and DF attain feature reduction by
dropping words from the vocabulary
(\cite{Yang:1997:CSF:645526.657137},
\cite{Forman:2003:EES:944919.944974}).

We chose Naive Bayes as a classification method since it lends itself
well for all four approaches.  Firstly, it allows for a
straight-forward transfer from words to topics as will be shown below.
Secondly, it does \emph{not} mandate additional hyper parameter
settings such as \emph{Support Vector Machines} (SVMs), which would complicate
the comparison and potentially incur bias.  Moreover, approaches
relying on a TF-IDF embedding (such as Roccio or SVM in
\cite{Joachims1998}) are problematic with regard to LDA because DF
and IDF are undefined for topics.

Note that our goal is \emph{not} to show that topic models can
generally reduce the word feature space without (much) loss of
classification accuracy.  This has already been demonstrated in
\cite{blei:2003:lda:944919.944937}.  Instead, \emph{we focus on the
  relative performance of the four feature reduction techniques}.
Including IG and DF allows for a direct comparison between topic
modeling and word selection methods.

Let $C = \{ c_1, \ldots, c_m \}$ be the set of classes for the
training documents $D$. We assume that the class assignments $l(d) \in C, d \in D$ are unique and known with regard to $D$.  We define $D_c$
as the subset of training documents belonging to class $c$, so $D_c = \{ d \in D | l(d) = c \}$.

When using topics, Naive Bayes determines the class of a test document $d_{test}$ via 
\[ argmax_{c \in C} \log p(c | d_{test}) \approx argmax_c \log (p(c) \cdot \prod_{t\in T} p(t|c)^{f_{d_{test}}(t)}).\]
with $p(c)$ estimated from by means of $p(c) \approx |D_c| / |D|$.

Regarding Topic Grouper, $f_{d_{test}}(t)$ and $p(t|c)$ can be estimated via the topic-word assignments $t(w)$ from Section \ref{basics}.  In total, this results in the following classification formula for Topic Grouper:
\[ argmax_c \log (|D_c| / |D|) + \sum_{t\in T} f_{d_{test}}(t) \cdot \log ((1 + \sum_{d \in D_c} f_d(t)) / (n + \sum_{d \in D_c} |d|)).\]
The ``$1 +$'' and ``$n + $'' in the second $\log$-expression form a standard Lidstone smoothing accounting for potential zero probabilities of the estimated $p(t|c)$. Other than that, its practical effects are effect negligible.

For best possible results under LDA, we estimate $f_{d_{test}}(t) \approx |{d_{test}}| \cdot p(t|d_{test}) $. In order to compute $p(t|d_{test})$ accurately, we resort to
the so-called fold-in method: A topic-word assignment $z_i$ is sampled for every word occurrence $w_i$ in $d_{test}$ using Gibbs sampling. This involves the use of the underlying topic model $\Phi$ and leads to a respective topic assignment vector $\textbf{z}$ of length $|d_{test}|$. More details on this sampling method can be found in Section 3 of \cite{wallach-etal:2009:icml}.
The procedure is repeated $S$ times leading to $S$ vectors $\textbf{z}^{(s)}$. Together, these results form the basis of
\[ p(t|d_{test}) \approx 1/S \cdot \sum_{s=1}^S 1/ |d_{test}| \sum_{i=1}^{|d_{test}|} \delta_{\textbf{z}_i^{(s)},t}.\]
Moreover, we estimate $p(t|c) \approx (\sum_{d \in D_c} p(t|d) \cdot |d|) / \sum_{d \in D_c} |d|$. In this case, an approximation of $p(t|d)$ is known from running LDA on the training documents.

As known from \cite{Joachims1998} Naive Bayes is robust against a large number of features, i.e. words, and performs best without any feature reduction.
So, one cannot hope for increasing classification accuracy but only for little loss in accuracy when transferring to an ever smaller number of topics.
The results are also a rough indicator of how well topics coincide with a human classification scheme: If topics tended to cover many words across classes, the probabilities $p(t|c)$ would be less peaked and Naive Bayes' classification accuracy would suffer (more). 

We work with two popular datasets, namely ``Reuters 21578'' and ``Twenty News Groups'':
\begin{itemize}
\item Reuters 21578\footnote{See \url{http://www.daviddlewis.com/resources/testcollections/reuters21578/} (cited 2018-03-04).} is text collection of business news in English with more than 120 class labels, most of them rarely occurring, and 21,578 (partly unlabeled) documents. We chose the ten most frequent labels and kept all documents with exactly one class label. Moreover, we applied the so-called ModApte split, leading to 7,142 documents for training and 2,513 for test. We performed (Porter) stemming and kept every stem that occurs at least three times in the training data. This way, we ended up with a training vocabulary of 9,567 stemmed words excluding stop words.

\item Twenty News Groups is a collection of newsgroup messages covering twenty areas of social discussion. We used a reworked version of the collection consisting of 18,846 documents each belonging to just one class.\footnote{See \url{http://qwone.com/~jason/20Newsgroups/  (cited 2018-03-04)}.}
We applied a random split into training and test documents based on a 75\% to 25\% ratio. Again, we performed (Porter) stemming and kept every stem that occurs at least five times in the dataset. Moreover, we removed all tokens containing non-alphabetical characters or being shorter than three characters. This way, we ended up with a training vocabulary of 25,826 stemmed words excluding stop words.

\end{itemize}

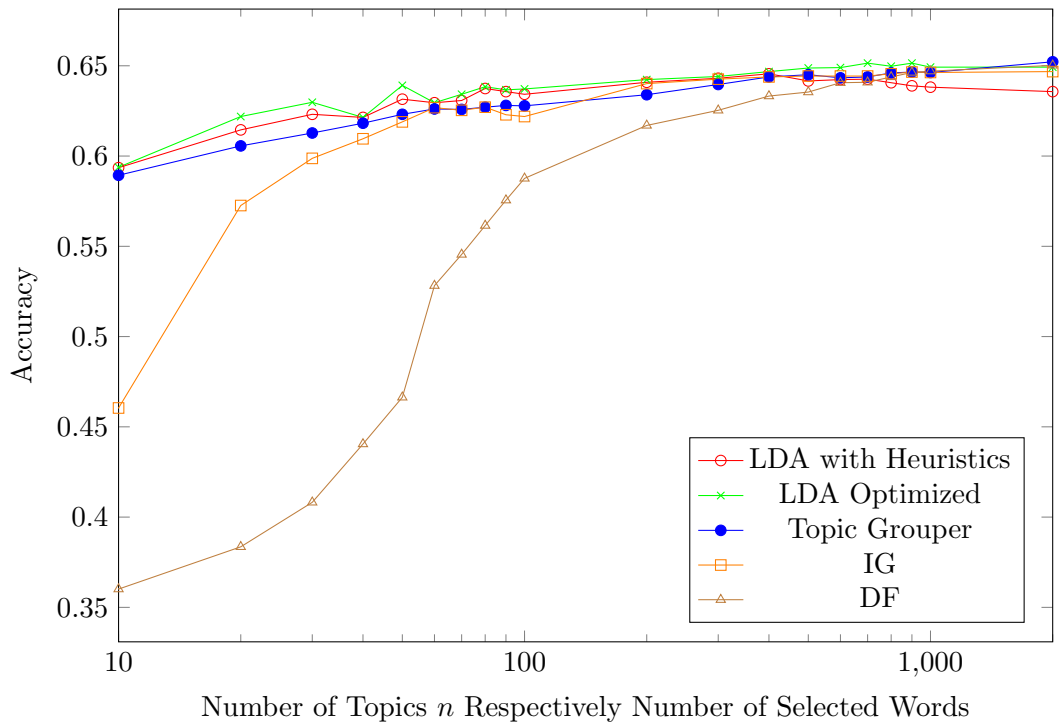
\begin{figure}
\begin{center}
\begin{tikzpicture} 
\begin{axis}[
  xmode=log,
  log ticks with fixed point,
  xlabel=Number of Topics $n$ Respectively Number of Selected Words,
	xmin=10,
	xmax=2000,
  ylabel=Accuracy,
	width=14cm,
	height=10cm,
	legend pos=south east
]
\addplot[color=red, mark=o] table [x=topics, y=microAvg, col sep=semicolon] {./csv/classification/ReutersLDAClassificationExperiment.csv};
\addlegendentry{LDA with Heuristics}
\addplot[color=green, mark=x] table [x=topics, y=microAvg, col sep=semicolon] {./csv/classification/ReutersLDAClassificationExperimentOpt.csv};
\addlegendentry{LDA Optimized}
\addplot[color=blue, mark=*] table [x=topics, y=microAvg, col sep=semicolon] {./csv/classification/ReutersTGNaiveBayesExperiment.csv};
\addlegendentry{Topic Grouper}
\addplot[color=orange, mark=square] table [x=topics, y=microAvg, col sep=semicolon] {./csv/classification/ReutersVocabIGClassificationExperiment.csv};
\addlegendentry{IG}
\addplot[color=brown, mark=triangle] table [x=topics, y=microAvg, col sep=semicolon] {./csv/classification/ReutersVocabDFClassificationExperiment.csv};
\addlegendentry{DF}
\end{axis}
\end{tikzpicture} 
\caption{Micro Averaged Classification Accuracy of Naive Bayes on Reuters 21578 Depending on the Log Scaled Number of Features}
\label{reutersclassresult}
\end{center}
\end{figure}

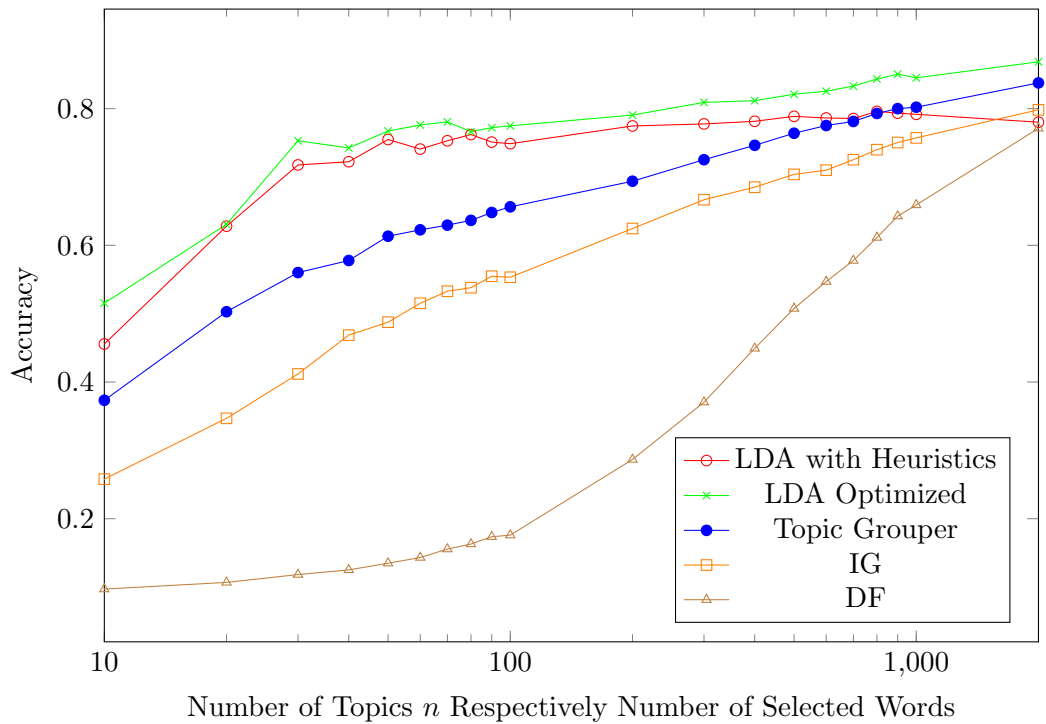
\begin{figure}
\begin{center}
\begin{tikzpicture} 
\begin{axis}[
  xmode=log,
  log ticks with fixed point,
  xlabel=Number of Topics $n$ Respectively Number of Selected Words,
	xmin=10,
	xmax=2000,
  ylabel=Accuracy,
	width=14cm,
	height=10cm,
	legend pos=south east
]
\addplot[color=red, mark=o] table [x=topics, y=microAvg, col sep=semicolon] {./csv/classification/TwentyNGLDAClassificationExperiment.csv};
\addlegendentry{LDA with Heuristics}
\addplot[color=green, mark=x] table [x=topics, y=microAvg, col sep=semicolon] {./csv/classification/TwentyNGLDAClassificationExperimentOpt.csv};
\addlegendentry{LDA Optimized}
\addplot[color=blue, mark=*] table [x=topics, y=microAvg, col sep=semicolon] {./csv/classification/TwentyNGTGNaiveBayesExperiment.csv};
\addlegendentry{Topic Grouper}
\addplot[color=orange, mark=square] table [x=topics, y=microAvg, col sep=semicolon] {./csv/classification/TwentyNGVocabIGClassificationExperiment.csv};
\addlegendentry{IG}
\addplot[color=brown, mark=triangle] table [x=topics, y=microAvg, col sep=semicolon] {./csv/classification/TwentyNGVocabDFClassificationExperiment.csv};
\addlegendentry{DF}
\end{axis}
\end{tikzpicture} 
\caption{Micro Averaged Classification Accuracy of Naive Bayes on Twenty New Groups Depending on the Log Scaled Number of Features}
\label{twentyngclassresult}
\end{center}
\end{figure}

Figures \ref{reutersclassresult} and \ref{twentyngclassresult} present
classification accuracy as a function of the number topics
or words, respectively, using micro averaging. Our findings confirm the
impressive abilities of LDA for feature reduction as reported in
\cite{blei:2003:lda:944919.944937} when applying hyper parameter
optimization.  Beyond 700 topics, the heuristic setting degrades LDA's
performance.  In accordance with \cite{Yang:1997:CSF:645526.657137}
and \cite{Forman:2003:EES:944919.944974}, the results confirm that IG
performs better than DF. The performance of Topic Grouper depends on
the dataset and ranges below ``LDA Optimized'' but considerably above
IG in Figure \ref{twentyngclassresult} whereas in Figure
\ref{reutersclassresult} ``LDA Optimized'', IG and Topic Grouper are close above 200 topics or words, respectively.

When applying topic modeling this way, an important point to consider is the computational overhead for model generation but also the feature reduction overhead for new documents at classification time:
Once a Topic Grouper model is built, its use for feature reduction incurs minimal overhead: I.e., a word from a test document $d_{test}$ can be reduced in constant time via the topic-word assignment $t(w)$.
Thus the total feature reduction cost for a test document remains on the order of $|d_{test}|$.
In contrast, LDA requires the relatively complex fold-in computation of $p(t|d_{test})$ which is on the order of $S \cdot |d_{test}| \cdot n$ for a test document.

Model generation for LDA tends to become computationally expensive, when the number of topics is high because it depends linearly on $|T|$.
We experienced this when producing the results above about $n \geq 500$. In comparison, Topic Grouper's computation time remained moderate even for the Twenty New Groups dataset with $|V| > 25,000$.
As noted before, Topic Grouper assesses all values for $n$ between $|V|$ and one within an single run.
The latter \emph{allows to adjust the degree of feature reduction in hindsight} without the need for topic model recomputations.

We believe that this favorable combination of qualities places Topic Grouper as a promising alternative to IG and DG with actual practical relevance.

\subsection{Model Visualization and Inspection}
\label{viz}

Topic Grouper returns hierarchical topic models by design.
The hierarchy of topics may be explored interactively assuming that larger topics form a kind of semantic abstraction of contained smaller topics. Much as under LDA, the meaning of a topic may be represented by its top-most frequent words on every containment level. Analyzing results this way may give users additional insight into the nature of a document collection's inherent topics.

\begin{figure}
\center{\includegraphics[width=15cm]{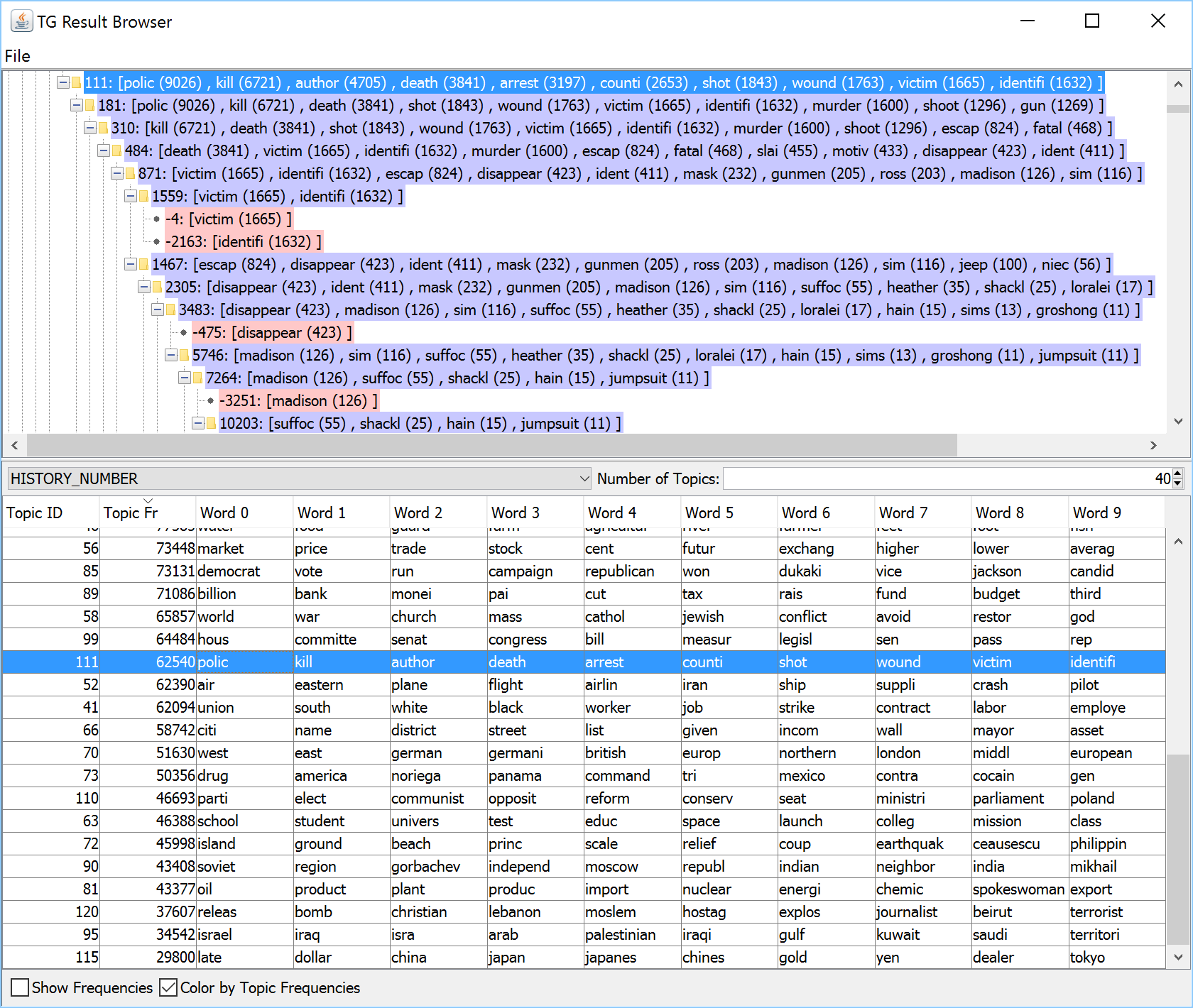}}
\caption{Screen Shot of a Simple Tool to Explore the Containment Hierarchy of Topic Models Produced by Topic Grouper}
\label{screenshot}
\end{figure}

Figure \ref{screenshot} shows a screen shot of a simple tool that we built for this purpose. 
The upper half of the window allows for exploring the containment structure of topics via a hierarchy of folders.
The lower half of the window displays a \textit{flat topic view}, which is a list of topics $T(n)$ as they occur together during a run of Topic Grouper according to Section \ref{basics}. The number $n$ can be changed interactively causing an instant update of the displayed topic list. Each topic from the list is displayed in one table row with the ten most frequent words included. A click on a table row selects the corresponding hierarchy node in the upper half of the window.
The depicted model in Figure \ref{screenshot} is Topic Grouper's result on the AP Corpus dataset from Section \ref{realworlddata}.

\begin{figure}
\center{\includegraphics[width=15cm]{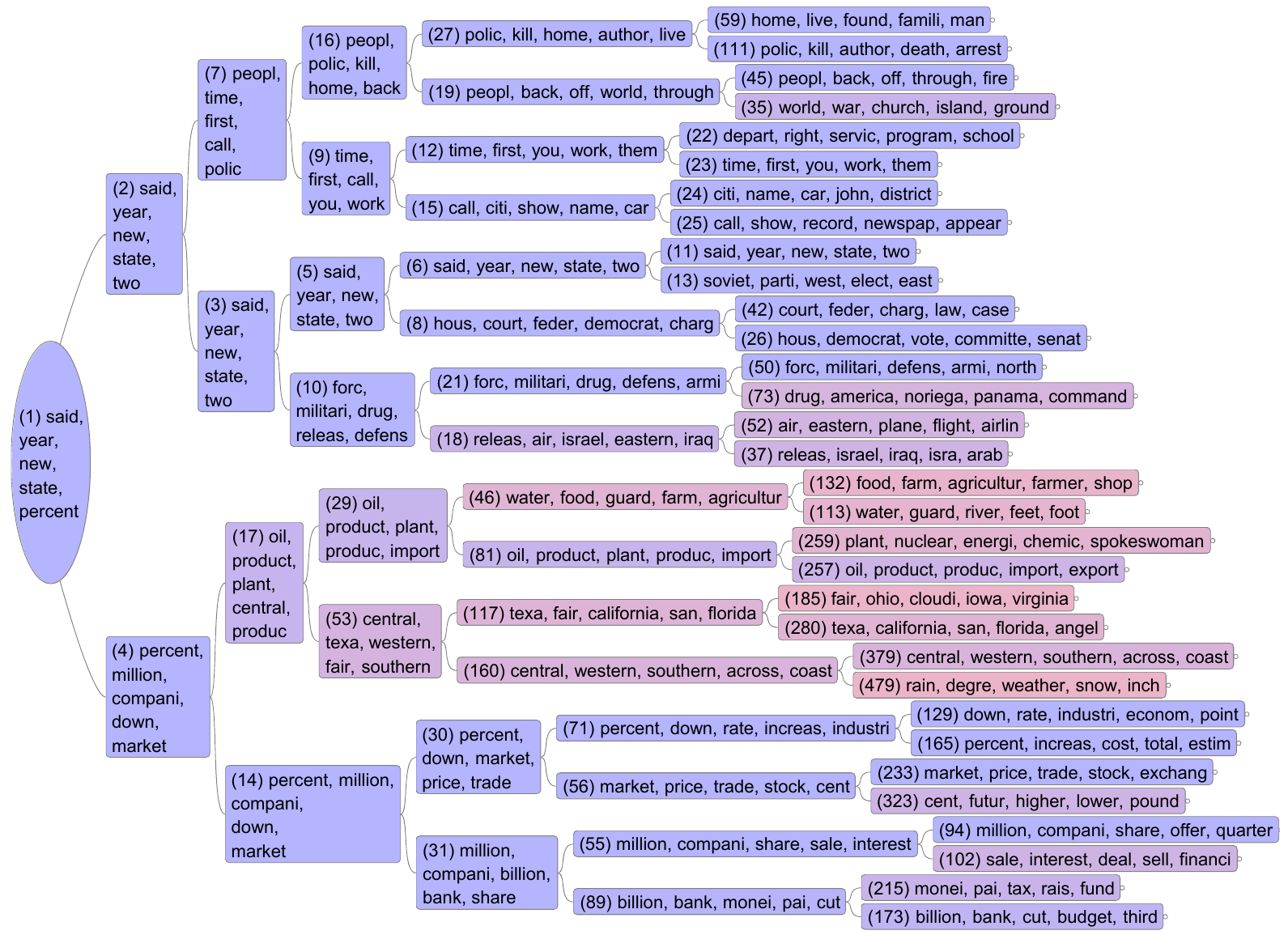}}
\caption{Mind Map Diagram as a result of Topic Grouper on the AP Corpus Extract}
\label{mindmap2}
\end{figure}

To reflect the containment hierarchy of topics, we also created tree diagrams using the mind map tool FreeMind.\footnote{See
\url{http://freemind.sourceforge.net}.} Topics are represented as nodes and for reference, they are identified by the number $n$ under which they were generated.
Figure \ref{mindmap2} presents a corresponding mind map for the AP Corpus dataset from Section \ref{realworlddata}. All nodes below level six are collapsed in order to deal with limited presentation space. A node contains the five most frequent words of a respective topic. More frequent topics are shaded in blue (as they tend to collect low content words and stop words), whereas less frequent word sets are shaded in red.

The contents of the tree may be interpreted is as follows: The root forks into node (4) covering economy and weather as well as node (2) covering other topics and function words. Function words are mainly gathered along the path (1)/(2)/(3)/(6)/(11) and the sub-path (9)/(12)/(23).
Node (4) forks into financial topics (14) and topics covering production and weather (17). Node (53) is on weather and potentially different weather regions. Node (46) covers agriculture and water supply whereas node (81) focuses on energy.
Regarding node (14), we suspect that stock trading in (30) is separated from general banking and aquisitions in (31). Other topics in the tree seem equally coherent such as ``home and family'' (59), ``public media'' (25), ``jurisdiction and law'' 
(42), ``military and defense'' (50) and so forth. We find that such interpreted topics often meet the idea of being more general towards the root and more specific towards the leaves. However mixed topics also arise such as topic (21) combining ``drug trafficking'' in (73) with ``military and defense'' in (50).

Table \ref{topiclist} lists topics from $T(40)$ for the AP Corpus dataset. To save presentation space only every second topic in order of frequency is shown: Topics 47 and 69 gather function words and therefore have high frequency. Most topics seem conclusive but obviously, a more objective coherency analysis would be necessary. A corresponding study with human judges may follow the approach in \cite{DBLP:conf/nips/ChangBGWB09} but is beyond the scope of this article.


\begin{figure}
\begin{center}
\small
\begin{tabular}{|c|c|ccccccc|}\hline 
$n$ & $f(t)$ & \multicolumn{7}{c|}{\bfseries Top Seven Words per Topic $t$} \\\hline
\begin{filecontents*}{aplargetopics2nd.csv}
topicid;topicfr;worda;wordb;wordc;wordd;worde;wordf;wordg;wordh;wordi;wordj
47;538739;year;new;two;dai;week;three;month;sinc;fridai;mai
40;305812;state;govern;nation;unit;american;includ;countri;group;member;leader
69;281349;said;report;offici;sai;befor;against;told;did;public;sever
42;176138;court;feder;charg;law;case;rule;order;investig;judg;attornei
71;119423;percent;down;rate;increas;industri;econom;point;board;cost;total
51;115641;presid;bush;plan;meet;talk;administr;propos;both;reagan;secretari
59;112332;home;live;found;famili;man;children;life;women;men;known
67;96161;commun;visit;miss;travel;becam;histori;art;room;often;trip
49;89151;call;show;newspap;appear;televis;radio;publish;station;telephon;daili
74;82919;john;william;robert;richard;paul;wait;king;brown;thoma;smith
46;77385;water;food;guard;farm;agricultur;river;farmer;feet;foot;fish
85;73131;democrat;vote;run;campaign;republican;won;dukaki;vice;jackson;candid
58;65857;world;war;church;mass;cathol;jewish;conflict;avoid;restor;god
111;62540;polic;kill;author;death;arrest;counti;shot;wound;victim;identifi
41;62094;union;south;white;black;worker;job;strike;contract;labor;employe
70;51630;west;east;german;germani;british;europ;northern;london;middl;european
110;46693;parti;elect;communist;opposit;reform;conserv;seat;ministri;parliament;poland
72;45998;island;ground;beach;princ;scale;relief;coup;earthquak;ceausescu;philippin
81;43377;oil;product;plant;produc;import;nuclear;energi;chemic;spokeswoman;export
95;34542;israel;iraq;isra;arab;palestinian;iraqi;gulf;kuwait;saudi;territori
\end{filecontents*}
\csvreader[separator=semicolon,head to column names,late after line=\\\hline]{aplargetopics2nd.csv}{}
{\topicid & \topicfr & \worda & \wordb & \wordc & \wordd & \worde & \wordf & \wordg}
\end{tabular}
\end{center}
\caption{Every Second Topic of $T(40)$ Sorted by Topic Frequency for the AP Corpus Dataset}
\label{topiclist}
\end{figure}

\section{Summary and Discussion} \label{sec:discussion}

We have presented Topic Grouper as a novel and complementary method in the field of probabilistic topic modeling based on agglomerative clustering:
Initial clusters or topics, respectively, each consist of one word from the vocabulary of the training corpus.
Clusters are joined on the basis of a simple probabilistic model assuming that each word belongs to exactly one topic. Thus, topics or clusters form a disjunctive partitioning of the vocabulary. 

After developing a related cluster distance $\Delta h$ we have adapted an existing clustering algorithm, EHAC, in order to compute related cluster trees as models.
Dendrogram cuts in the tree serve as flat topic views where a fixed number of topics may be chosen in the range of the vocabulary size $|V|$ and one.
The adapted clustering algorithm makes use of the dynamic programming principle leading to a time complexity in $O(|V|^2 \cdot |D|)$ and a space complexity in $O(|V|^2)$, where $|D|$ is the number of training documents.
Since memory consumption may be of an issue, we devised an additional algorithm, MEHAC, with an \emph{expected} time complexity in $O(|V|^2 \cdot |D|)$ but space complexity only on the order of $|V|$.

Using simple synthetic datasets, where each word belongs to just one original topic, we examined some basic qualities of topic modeling methods:
Topic Grouper manages to recover related original topics at low error rate even when their a-priori probabilities are rather unbalanced.
pLSI fails under these conditions. LDA is able to recover the original topics but only if its vectorial hyper parameter $\alpha \textbf{m}$ is adjusted accordingly.

Regarding various real world datasets, Topic Grouper's predictive performance matched or surpassed LDA with Heuristics at larger topic numbers but was still dominated by LDA Optimized, where only the latter includes an optimization for the LDA-specific hyper parameters $\alpha \textbf{m}$ and $\beta$ but the former applies a commonly used heuristic for them.
The results also suggest that Topic Grouper performs the better the less polysemy there is in the vocabulary. This is consistent with Topic Grouper's simplifying topic models. 
It makes the approach appealing, for instance, for shopping basket analysis where articles stand for themselves: Related models may then aid in forming sales-driven catalog structures or layouts of product assortments since
in both cases, a clear-cut to decision on where to place an article is customary.

We also investigated Topic Grouper as a means for feature reduction in the field of supervised text classification: The results suggest that it outperforms standard techniques in the field such as 
Information Gain (IG) and Document Frequency (DF) but is dominated by LDA Optimized. However, LDA incurs a considerable runtime overhead at classification time, where Topic Grouper does not.
Also Topic Grouper allows for a dynamic change of the number of topics after training, whereas LDA would require retraining.

Based on a corpus of news wire articles (AP Corpus) we showed how a tree model produced by Topic Grouper may be visualized and explored interactively. The presented corpus results exhibit the descriptive qualities of such
deep tree models as well as the potential of related drill downs from more general to more specific topics. Alternatively, flat views of an arbitrary number of topics between $|V|$ and one may be derived instantly from the generated model. Although this is a subjective impression, we found corresponding topics to be conclusive and coherent in both tree views and flat topic views. Obviously, this assessment demands a more objective study to follow, potentially in similarity to \cite{DBLP:conf/nips/ChangBGWB09}, \cite{Newman:2010:AET:1857999.1858011} or \cite{lau-newman-baldwin:2014:eacl}. 

For all text corpora we found that Topic Grouper tends to push stop words and or function words into separate topics. Therefore, it can do without stop word or function word filtering as a preprocessing step.

The practical performance of our straight forward implementation ranged between several minutes to several hours for larger datasets of this report and substantiated the theoretical complexity. 
\emph{A simple and effective means to increase time and space performance is obviously to reduce the vocabulary size $|V|$, e.g. by keeping only a few thousand topmost frequency words from the dataset.}
The approach is well in line with the standard practice to focus on high probability words or in case of Topic Grouper, on high frequency words, when displaying and inspecting a topic model's topic-word distributions.

In conclusion, we see Topic Grouper as a complementary approach in the tool set of topic modeling methods with a unique mix of pros and cons. The tree-based model also offering flat topic views is an important asset. It allows for deep tree structures to be produced even on small-sized datasets.
Another benefit is the method's simplicity and that it requires no configuration or hyper parametrization and no stop word filtering. 
The fact that each word is in exactly one topic is a considerable limitation and falls short for 
polysemic words and for words applied in multiple topical contexts. Nevertheless, we found actual 
topic models for text corpora to be conclusive as reported in Section \ref{viz}.  
In some cases, a clear-cut decision on where to place words may even be in accordance with an analysts interests---we mentioned examples regarding shopping basket analysis.

The results of this paper can all be reproduced via a prototypical Java library named ``TopicGrouperJ'' published on GitHub.\footnote{See \url{https://github.com/pfeiferd/TopicGrouperJ}.}
The library features implementations of the corresponding algorithms MEHAC and EHAC. Amongst other things, 
it contains an LDA Gibbs Sampler with options for hyper parameter optimization and an implementation to compute perplexity as discussed in Section \ref{perplexity}. The code to regenerate any result file of the above-described experiments is also available.

\section{Future Work} \label{sec:conclusion}

Future research directions may include the \emph{parallelization} of the Topic Grouper algorithms MEHAC and EHAC along with other computational optimizations. Note that the parts of EHAC affecting data structure updates after joining two topics are straight forward to parallelize.

An important concern for further work is \emph{model smoothing}, i.e. on how to relax the constraint of each word being in exactly one topic: Regarding flat topic views, we experimented with a combination of Topic Grouper and LDA, where LDA acts as post-processing step. To do so, a flat topic view $T$ from Topic Grouper is used to set the LDA hyper parameter $\beta$ then formed as a matrix in $\Re^{|V| \times |T|}$ where each column corresponds to a designated topic $t \in T$. Higher values for a matrix element in $t$'s column will be given if a corresponding relation $w \in t$ holds. A resulting LDA model $\Theta$ will then be close to the original topics $T$ from Topic Grouper but allows for other words to be included to a certain degree in each distribution $\Theta_t$. Compiling related experimental results is work in progress. Alternatively, topics produced by Topic Grouper may provide useful initialization values for an EM procedure under pLSI.

Another line of research may be the early \emph{detection of polysemic words} $w$ in order to address them in a special manner during the clustering process. I.e., if $\Delta h(\{w\}, s)$ and $\Delta h(\{w\}, t)$ according to Equation \ref{eq:deltah} are  similar and high, then the topics $s$ and $t$ are both good join candidates for $\{w\}$. This may trigger a special treatment of $w$.

We have already mentioned the need to substantiate model quality via \emph{extrinsic evaluation} methods as described to in Section \ref{perplexity}.

Our tool from Section \ref{viz} allows for just a basic exploration of learned tree models.
A more sophisticated system may include complementary visualization methods and the aforementioned smoothing procedures for flat topic views. Also, navigational links from topics to their underlying documents are to be included.
\emph{Tool support for the exploration of document collections} is an ongoing area of research and many solutions have been suggested---several of them exploiting LDA topic models (e.g., \cite{conf/icwsm/ChaneyB12}, \cite{Gretarsson:2012:TVA:2089094.2089099}, \cite{lee-etal:2012:compgraphfor} or \cite{sievert2014ldavis}). Corresponding insights and concepts should be considered and potentially adapted when leveraging results from Topic Grouper. In this context, a particular question is how take advantage of related tree models as opposed to the established use of flat topics. Note that topic trees are demanded by certain clientele: E.g., \cite{brehmer-etal:2014:tvcg} stress their importance when reporting about the Overview system -- a successful document analysis tool developed with a focus on investigative journalism.


\cite{Wei-Croft:2006:SIGIR} have employed LDA to \emph{improve document ranking models} for ad-hoc document retrieval. Their approach may be adapted to use models from Topic Grouper instead. 
The efficiency of determining $t(w)$ and $p(t|d_{test})$ under Topic Grouper may generally be useful to improve retrieval results: E.g., \emph{query expansion} may be performed on 
the basis of small topics $t$ containing all or most of the entered search terms $w$. In this regard, best matching topics may be chosen from the entire topic tree -- not just a flat view of topics.

Finally, topic modeling has been applied to the field of \emph{recommender systems} (e.g. see \cite{Wang:2011:CTM:2020408.2020480, hu-hall-attenberg:2014:kdd}). Consequently, it might be interesting to assess the potential of Topic Grouper for this purpose as it produces even very small topics and may therefore, play a similar role for recommendation as the Apriori method (\cite{Sandvig:2007:RCR:1297231.1297249}).




\appendix

\section{Perplexity on Datasets According to \cite{tan-ou:2010:iscslp}}
\label{perplexity_syn}

\begin{figure}
\begin{center}
\begin{tikzpicture} 
\begin{axis}[
  xlabel=$\alpha \textbf{m}_1$,
  ylabel=Perplexity,
	width=10cm,
	height=9cm
]
\addplot[smooth,color=black,mark=|] table [x=alpha1, y=perplexity, col sep=semicolon] {./csv/twc/TWCUnigramPerplexityChangeAlphaExp.csv};
\addlegendentry{Unigram 1}

\addplot[smooth,color=gray] table [x=alpha1, y=perplexity, col sep=semicolon] {./csv/twc/TWCUnigramPerplexityChangeAlphaExp2.csv};
\addlegendentry{Unigram 2}

\addplot[smooth,color=brown,mark=triangle*,error bars/.cd, y dir = both, y explicit] table [x=alpha1, y=perplexityLRAvg, y error=perplexityLRStdDev, col sep=semicolon] {./csv/twc/TWCPLSAPerplexityChangeAlphaExp.csv};
\addlegendentry{pLSI 1}

\addplot[smooth, color=red, mark=o,error bars/.cd, y dir = both, y explicit] table [x=alpha1, y=perplexityLRAvg, y error=perplexityLRStdDev, col sep=semicolon] {./csv/twc/TWCLDAPerplexityChangeAlphaExp.csv};
\addlegendentry{LDA 1}

\addplot[smooth, color=orange, mark=x,error bars/.cd, y dir = both, y explicit] table [x=alpha1, y=perplexityLRAvg, y error=perplexityLRStdDev, col sep=semicolon] {./csv/twc/TWCLDAPerplexityChangeAlphaExp2.csv};
\addlegendentry{LDA 2}

\addplot[smooth,color=blue,mark=*] table [x=alpha1, y=perplexity, col sep=semicolon] {./csv/twc/TWCTGPerplexityChangeAlphaExp.csv};
\addlegendentry{Topic Grouper 1}

\addplot[smooth,color=violet,mark=triangle] table [x=alpha1, y=perplexity, col sep=semicolon] {./csv/twc/TWCTGPerplexityChangeAlphaExp2.csv};
\addlegendentry{Topic Grouper 2}

\addplot[smooth,color=green,mark=square*] table [x=alpha1, y=perplexity, col sep=semicolon] {./csv/twc/TWCPerfectPerplexityChangeAlphaExp.csv};
\addlegendentry{Perfect 1}

\addplot[smooth,color=lime,mark=square] table [x=alpha1, y=perplexity, col sep=semicolon] {./csv/twc/TWCPerfectPerplexityChangeAlphaExp2.csv};
\addlegendentry{Perfect 2}

\end{axis}
\end{tikzpicture}
\caption{Perplexity Depending on $\alpha \textbf{m}_1$ for Two Datasets Generated According to \cite{tan-ou:2010:iscslp}}
\label{ldasympresult1}
\end{center}
\end{figure}
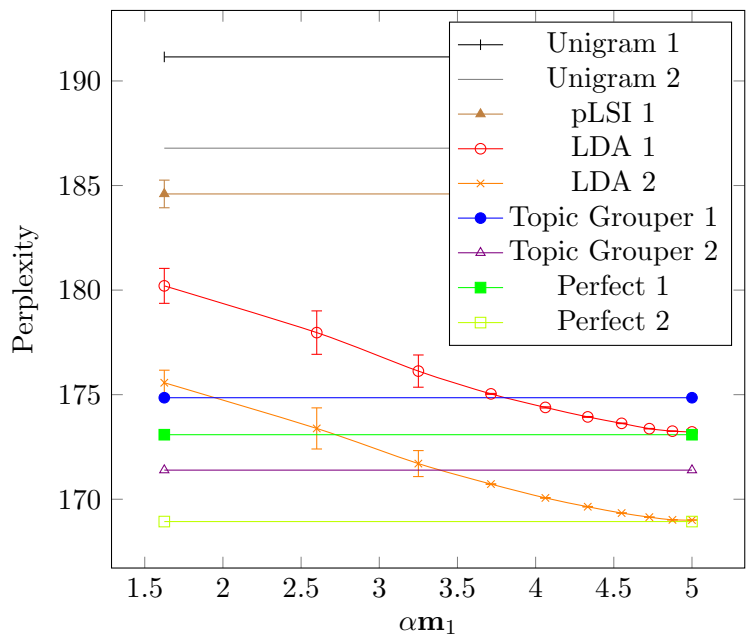

Figure \ref{ldasympresult1} depicts the same scenario as Figure
\ref{ldasymresult} but with a focus on perplexity (as computed according to Section \ref{realworlddata}). Interestingly, \emph{it
suggests that perplexity is not necessarily a good ``substitute
measure'' for error rate}:
\begin{itemize}
\item Although error rate and perplexity are correlated the relative difference in perplexity with regard to low
  and high error rates is small.
\item The base values for perplexity differ considerably for the two
  datasets (although they have been generated under the same data
  generator based on identical hyper parameters).
\item The low error rate of Topic Grouper does not transfer to a
  correspondingly low perplexity, while the one of LDA does.
\end{itemize}

\section{Memory Efficient Agglomerative Clustering for Topic Grouper (MEHAC)}
\label{mehac}

This algorithm for Topic Grouper offers a small memory footprint on the order of $|V|$ and expected time complexity in $O(|V|^2 \cdot |D|)$.
For every topic $s \in T$ it keeps only the best possible join partner $t \in T, t \neq s$ according  to $\Delta h$ from Equation \ref{eq:deltah}. All respective pairs $(s,t)$ are stored in a single priority queue sorted by descending $\Delta h(s,t)$ values.

The code includes an important optimization in line 59, which considerably improves its practical performance, but does not affect its order of complexity. It will be discussed at the end of this section.

\begin{enumerate}
\item The process starts by computing the one-word topics and the best possible join partner for every topic $s$ with $|s| = 1$ (lines 15 to 32). (The runtime of the lines 23 to 32 can be halved by exploiting the symmetry $h(t \cup s) = h(s \cup t)$. For simplicity, this is omitted in the code.)

\item At each step $i > 0$ the algorithm picks an $(s,t)$ from the top of the priority queue. If $(s,t)$ is valid (concerning validity, see point 4), it joins $s$ and $t$ and updates related data structures (line 39 to 47). Moreover, it recomputes the best possible join partner $q$ for $s \cup t$ with $n - 2$ potential join partners available (line 48). This includes inserting the pair $(s \cup t, q)$ according to its $\Delta h$-value.

\item  Next, it adjusts the best possible join partners for all topics $w \in T(n)$ with $w \neq s\cup t$ (line 49). It does so by checking whether $s \cup t$ is a better join partner for $w$ than $v$. If so, then $(w,v)$ is removed from list and $(w, s \cup t)$ is inserted as unmarked on the basis if its $\Delta h$-value.

\item If $w$'s best possible join partner so far was $s$ or $t$, then the algorithm defers the computation of $w$'s new best possible join partner by marking $w$'s current entry in the queue and setting $v$ to $null$ (lines 58 to 61). The position of the corresponding pair in the queue remains unchanged, though (line 59). The mark ensures that the pair $(w, null)$ will be considered \emph{invalid} in case it is picked as $(s,t)$ from the top of the queue according point 2) (line 35). If this happens, then $(s,t)$ will be removed but $s$ and $t$ will not be joined. Instead, the best possible join partner $q$ for $s$ will be recomputed and $(s,q)$ will be inserted into the priority queue according to its $\Delta h$-value (line 36).
\end{enumerate}

According to Section \ref{initialization}, the complexity of point 1) is on the order of $|V|^2 \cdot |D|$.

The computational cost of a step $i > 0$ is on the order of $|V| \cdot |D|$: To determine the new best possible join partner for $s \cup t$, every other topic $r$ in $T(n)$ with $r \neq s \cup t$ must be considered and $\Delta h((s \cup t), r)$ must be computed by iterating over $D$ (lines 68 and 73). Similarly, checking whether $s \cup t$ is potentially a better join partner for any $w$ with $w \neq s\cup t$ is on order of $|V| \cdot |D|$ (lines 49 and 60). Adding up over all steps $i$, the complexity remains on the order of $|V|^2 \cdot |D|$.

One might wonder whether the complexity is worsened by the number of invalid pairs $(s,t)$ getting picked from the top list (line 36): 
But note that the update in line 36 represents a deferred computation that could have as well been done right away in line 59: If the update was not deferred, the probability of the condition from line 57 to be true is expected to be less than $c / (|T| - 2)$ where $c \geq 2$ is a constant depending on the data. Given so, the expected number of respective non-deferred updates in the loop from line 49 would be less than $(|T| - 1) \cdot c / (|T| - 2) \in O(1)$.
We found that by deferring related updates, a substantial fraction of it can be avoided and thus performance be improved, since an update must only be executed if a corresponding 
tuple $(w, null)$ appears at the top of queue.

We also tested the frequency of related updates for the deferring case with regard to various datasets including all the ones from Section~\ref{sec:evaluation}. The results confirmed that the frequency remains in the order $|V|$. As the computation of the procedure call from line 36 is in the order of $|V| \cdot |D|$, the total order of complexity of the algorithm remains in $O(|V|^2 \cdot |D|)$.

Another point to mention is that for a marked and thus invalid pair $(w,null)$, its position in the sorted list remains ``optimistic'', i.e., rather too close to the top of queue (line 59). Therefore, $w$'s join partner will not be recomputed too late and no other pair of join partners will accidentally be given preference to be joined.

The algorithm can be readily adjusted for other agglomerative clustering tasks, while keeping its low memory footprint.

\lstset{numbers=left,morekeywords={new, foreach, var, foreach, procedure, print, insert, remove, add, null, while, true, false, clear},basicstyle=\tiny,escapeinside={(*}{*)}}
\begin{lstlisting}[mathescape=true,caption={Memory Efficient Agglormerative Clustering for Topic Grouper (MEHAC)},label=lst:mehac]
// (*\textbf{Input: $V, D, f_d(w)$ and $f(w)$ according to Section \ref{basics}}*)
// (*\textbf{Output: Relevant changes of T -- the current set of topics -- printed out.}*)

// (*\textbf{Global variables}*)
var T := $\emptyset$; // Current set of topics
// Priority queue of topic pairs (s,t), 
// where t is the best possible join partner for s. 
// It is sorted in descending order on the basis of $\Delta h(s,t)$:
var pq := new PriorityQueue();
// Map for parameters from Equation (*\ref{eq:efficienth}*), topics from T are used as keys:
var h[], f[], i[], fd[];
var $\Delta$h[]; // Map of topic pairs (s,t) to $\Delta$h-values

// (*\textbf{Initialization step $i$ = 0}*)
foreach $w \in V$ { // Filling T
  var t := { $w$ };
  insert t into T;
  h[t] := $h(t)$ according to Equation (*\ref{eq:init1}*)
  foreach $d \in D$ { fd[(t,d)] := $f_d(w)$; }
  f[t] := $f(w)$; i[t] := $f(w) \cdot \log f(w)$;
}
print T;
foreach s $\in$ T { // Computing initial best possible join partners
  var $\Delta$h := $-\infty$, t;
  foreach r $\in$ T with r $\neq$ s {
    var h := $h(r \cup s)$ according to Equation (*\ref{eq:init2}*);
    if (h - h[r] - h[s] > $\Delta$h) {
      $\Delta$h := h - h[r] - h[s]; t := r; }
  }
  add (s,t) to pq on the basis of $\Delta$h;
  $\Delta$h[(s,t)] := $\Delta$h;
}
// (*\textbf{Steps i $>$ 0 to join topics}*)
while (|T| > 1) {
  (s,t) := pq.poll;
  if (t = null) update_join_partner_for(s);
  else { 
    // (*\textbf{Step i to join topics s and t:}*)
    remove (t,?) from pq; // Remove all tuples from pq with t as first element.
    remove s from T; remove t from T;
    var u := s$\cup$t;
    insert u into T;
    print T;
    foreach $d \in D$ { fd[(u, d)] := fd[(s,d)] + fd[(t,d)]; clear fd[(s,d)], fd[(t,d)]; }
    f[u] := f[s] + f[t]; i[u] := i[s] + i[t]; 
    h[u] := $h(u)$ according to Equation (*\ref{eq:efficienth}*);
    clear h[s], h[t], f[s], f[t], i[s], i[t];
    update_join_partner_for(u);
    foreach (w,v) in pq with w $\neq$ u {
      var h := $h(w \cup u)$ according to Equation (*\ref{eq:efficienth}*);
      var $\Delta$h := h - h[u] - h[w];
      if ($\Delta$h > $\Delta$h[(w,v)]) {
        remove (w,v) from pq;
        add (w,u) to pq on the basis of $\Delta$h;			  
        $\Delta$h[(w,u)] := $\Delta$h;
      }
      else if (v = s or v = t) {
        remove (w,v) from pq;
        add (w,null) to pq on the basis of $\Delta$h[(w,v)];			  
      }
    }
  }
}

// (*\textbf{To find the best possible join partner r from T for topic u}*)
procedure update_join_partner_for(var s) {
  var $\Delta$h := -$\infty$; var t;
  foreach r $\in$ T with r $\neq$ s {
    var h := $h(r \cup s)$ according to Equation (*\ref{eq:efficienth}*);
    if (h - h[r] - h[s] > $\Delta$h) {
      $\Delta$h := h - h[r] - h[s]; t := r; }
  }
  add (s,t) to pq on the basis of $\Delta$h;
  $\Delta$h[(s,t)] := $\Delta$h;
}
\end{lstlisting}

\section{Practical Performance}
\label{pperformance}

This section gives a brief impression of the runtime performance of Topic Grouper.
We measured related times using a Java implementation of both algorithms EHAC and MEHAC (see Section \ref{complexity} and Appendix \ref{mehac}). A regular virtual computing environment was used (Intel Xeon E5-2690 processor, 2.9 GHz, 96 GB RAM, Ubuntu Server 16, Java 1.8) and output was minimized to the essentials.

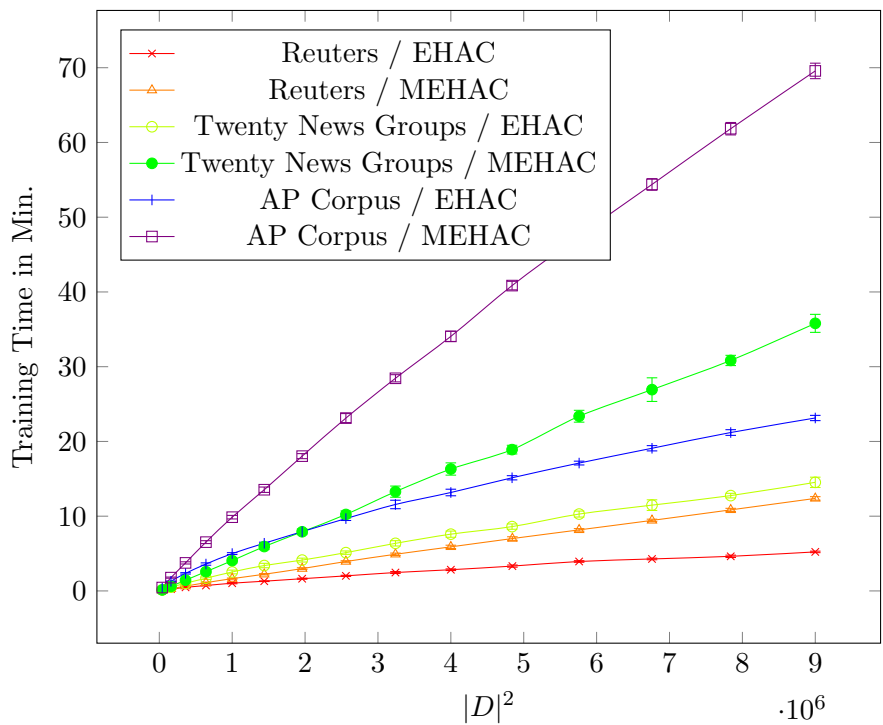
\begin{figure}
\begin{center}
\begin{tikzpicture} 
\begin{axis}[
  legend pos=north west,
  xlabel=$|D|^2$,
  ylabel=Training Time in Min.,
	width=12cm,
	height=10cm
]
\addplot[smooth, color=red, mark=x,error bars/.cd, y dir = both, y explicit] table [x expr={\thisrow{avgDocs} * \thisrow{avgDocs}}, y expr={\thisrow{avgTimeSec} / 60}, y error expr={\thisrow{stdDevTimeSec} / 60},col sep=semicolon] {./csv/performance/ReutersTGPChangeDocsExp.csv};
\addlegendentry{Reuters / EHAC}

\addplot[smooth, color=orange, mark=triangle,error bars/.cd, y dir = both, y explicit] table [x expr={\thisrow{avgDocs} * \thisrow{avgDocs}}, y expr={\thisrow{avgTimeSec} / 60}, y error expr={\thisrow{stdDevTimeSec} / 60},col sep=semicolon] {./csv/performance/ReutersLowMemTGPChangeDocsExp.csv};
\addlegendentry{Reuters / MEHAC}

\addplot[smooth, color=lime, mark=o,error bars/.cd, y dir = both, y explicit] table [x expr={\thisrow{avgDocs} * \thisrow{avgDocs}}, y expr={\thisrow{avgTimeSec} / 60}, y error expr={\thisrow{stdDevTimeSec} / 60},col sep=semicolon] {./csv/performance/TwentyNGTGPChangeDocsExp.csv};
\addlegendentry{Twenty News Groups / EHAC}

\addplot[smooth, color=green, mark=*,error bars/.cd, y dir = both, y explicit] table [x expr={\thisrow{avgDocs} * \thisrow{avgDocs}}, y expr={\thisrow{avgTimeSec} / 60}, y error expr={\thisrow{stdDevTimeSec} / 60},col sep=semicolon] {./csv/performance/TwentyNGLowMemTGPChangeDocsExp.csv};
\addlegendentry{Twenty News Groups / MEHAC}

\addplot[smooth, color=blue, mark=|,error bars/.cd, y dir = both, y explicit] table [x expr={\thisrow{avgDocs} * \thisrow{avgDocs}}, y expr={\thisrow{avgTimeSec} / 60}, y error expr={\thisrow{stdDevTimeSec} / 60},col sep=semicolon] {./csv/performance/APTGPChangeDocsExp.csv};
\addlegendentry{AP Corpus / EHAC}

\addplot[smooth, color=violet, mark=square,error bars/.cd, y dir = both, y explicit] table [x expr={\thisrow{avgDocs} * \thisrow{avgDocs}}, y expr={\thisrow{avgTimeSec} / 60}, y error expr={\thisrow{stdDevTimeSec} / 60},col sep=semicolon] {./csv/performance/APLowMemTGPChangeDocsExp.csv};
\addlegendentry{AP Corpus / MEHAC}

\end{axis}
\end{tikzpicture} 
\caption{Training Times for Topic Grouper Depending on the Squared Number of Documents Randomly Drawn from a Respective Dataset}
\label{tgperformance}
\end{center}
\end{figure}

Figure \ref{tgperformance} shows runtimes in minutes with regard to text-based datasets from Section \ref{realworlddata} and \ref{featurered} depending on the \emph{squared} number of training documents $|D|^2$. Related documents were randomly drawn from the original training datasets. A data point is an average over 10 runs whereby the random subsample changed for each run. For each graph the number of documents ranged between 0 and 3000. The graphs substantiate that MEHAC is generally slower than EHAC. Also, the results are consistent with the simplified complexity $O(D^2)$ from Section \ref{complexity} incorporating Heaps' Law. However, we found that Heaps' Law, i.e. the relation $|V|^2 \sim |D|$, does not hold for the for retail datasets from Section \ref{realworlddata}.

To enable runtime prediction per dataset $i$, we determined a least squares regression function $\beta_i \cdot |V|^2 \cdot |D|$ for data points such as in Figure \ref{tgperformance} (not shown). The regression function has been chosen in concordance with the stated time complexity, $O(|V|^2 \cdot |D|)$, from Section \ref{complexity}. Table \ref{trainingtimes} shows the actual and estimated runtimes for several larger sized datasets used elsewhere in this report. The estimates are based on the regression parameters $\beta_i$, which may vary considerably not only between the two algorithms but also between the datasets. It is worth noting why the latter is the case: Regarding both EHAC and MEHAC Equations \ref{eq:efficienth}, \ref{eq:init1} and \ref{eq:init2} (as used in Listing \ref{lst:ehac}) refer to dataset specific subsets of documents incurring different runtimes of related loops. On top, MEHAC has a relevant dataset specific condition in Line 36 of Listing \ref{lst:mehac}. 

We shunned a direct performance comparison with LDA and its derivatives, since it is rather unclear what to compare against in detail.
There are many factors of LDA which have no counterpart in Topic Grouper such as hyper parameters and number of training iterations. Also, there are several approximization algorithms for LDA with different performance characteristics (\cite{Asuncion:2009:SIT:1795114.1795118}). 

\begin{figure}
\begin{center}
\small
\begin{tabular}{|c|r|r|r|r|r|r|r|}\hline
         &       &       &            & \multicolumn{2}{|c|}{\textbf{EHAC (Min.)}} & \multicolumn{2}{|c|}{\textbf{MEHAC (Min.)}}  \\
\textbf{Dataset} & $|V|$ & $|D|$ & \textbf{Avg.} $|d|$ & \textbf{Actual} & \textbf{Est.} & \textbf{Actual} & \textbf{Est.} \\
\hline
Online Retail & 3,464 & 17,086 & 154 & 6 & 8 & 37 & 32 \\
\hline
Ta Feng & 7,893 & 105,010 & 9 & 92 & 191 & 926 & 1,110 \\
\hline
AP Corpus & 25,047 & 17,989 & 260 & 529 & 887 & 1,660 & 2,527 \\
\hline
Reuters 21578 & 9,567 & 7,142 & 84 & 40 & 40 & 72 & 89 \\
\hline
Twenty New Groups & 25,826 & 14,129 & 150 & 583 & 623 & 1,030 & 1,450 \\
\hline
\end{tabular}
\end{center}
\caption{Size Characteristics and Topic Grouper Training Times in Minutes for Larger Datasets Used Elsewhere in this Report}
\label{trainingtimes}
\end{figure}

\newpage
\bibliography{literature}

\end{document}